%% file: ms.tex
\documentclass[reprint,prd,amsmath,amssymb,aps,superscriptaddress,twocolumn, floatfix, nofootinbib]{revtex4-1}
\pdfoutput=1
\usepackage{color, verbatim, ulem, graphicx, dcolumn, bm, hyperref, tabularx}

\newcommand{\apjl}{ApJLett}		
\newcommand{\apjs}{ApJS}		
\newcommand{\aap}{A\&A}			
\newcommand{\mnras}{MNRAS}		
\newcommand{\pasp}{Publications of the Astronomical Society of the Pacific}

\usepackage{numprint}
\usepackage{siunitx}
\usepackage{booktabs}

\newcommand\redsout{\bgroup\markoverwith{\textcolor{red}{\rule[0.5ex]{2pt}{0.4pt}}}\ULon}

\begin{document}

\title{Matter Power Spectrum Emulator for $f(R)$ Modified Gravity Cosmologies}

\author{Nesar Ramachandra}
\thanks{The two lead authors have contributed equally to this work.}
\affiliation{High Energy Physics Division, Argonne National Laboratory, Lemont, IL 60439, USA}
\author{Georgios Valogiannis}
\thanks{The two lead authors have contributed equally to this work.}
\affiliation{Department of Astronomy, Cornell University, Ithaca, NY 14853, USA}
\affiliation{Department of Physics, Harvard University, Cambridge, MA 02138, USA}
\author{Mustapha Ishak}
\affiliation{Physics Department, The University of Texas at Dallas, Richardson, Texas 75080, USA}
\author{Katrin Heitmann}
\affiliation{High Energy Physics Division, Argonne National Laboratory, Lemont, IL 60439, USA}
\collaboration{The LSST Dark Energy Science Collaboration}
\noaffiliation

\date{\today}

\begin{abstract}

Testing a subset of viable cosmological models beyond General Relativity (GR), with implications for cosmic acceleration and the Dark Energy associated with it, is within the reach of Rubin Observatory Legacy Survey of Space and Time (LSST) and a part of its endeavor. 
Deviations from GR-$w(z)$CDM models can manifest in the growth rate of structure and lensing, as well as in screening effects on non-linear scales.
We explore the constraining power of small-scale deviations predicted by the $f(R)$ Hu-Sawicki Modified Gravity (MG) candidate, by emulating this model with COLA (COmoving Lagrangian Acceleration) simulations. We present the experimental design, data generation, and interpolation schemes in cosmological parameters and across redshifts for the emulation of the boost in the power spectra due to Modified Gravity effects. Three preliminary applications of the emulator highlight the sensitivity to cosmological parameters, Fisher forecasting and Markov Chain Monte Carlo inference for a fiducial cosmology. This emulator will play an important role for future cosmological analysis handling the formidable amount of data expected from Rubin Observatory LSST. 

\end{abstract}

\maketitle

\section{Introduction}
\input{introduction_sec.tex}

\section{Training data and design}\label{Training}

\input{data_sec.tex}

\section{Emulator development}\label{Development}
\input{development_sec.tex}

\section{Emulator Accuracy and Verification}\label{results_sec}

\input{results_sec.tex}

\section{Emulator Preliminary Applications}\label{applications_sec}
\input{application_sec.tex}

\section{Conclusions}\label{conclusions_sec}
\input{conclusion_sec.tex}

\appendix

\section{Savitzky-Golay smoothing}\label{Smooth_app}
\input{smooth_appendix.tex}

\section{Gaussian Processes}\label{Gaussian_app}
\input{gp_appendix.tex}

\section{Parameters of COLA simulation}\label{sim_app}
\input{params_appendix.tex}

\acknowledgments

This paper has undergone internal review by the LSST Dark Energy Science Collaboration. The internal reviewers were Tessa Baker, Francois Lanusse and Danielle Leonard. The authors would like to thank them for useful feedbacks. 

The two lead authors, NR and GV have contributed equally to this work.
NR was involved in development and applications of the emulator, and contributed to the text of the paper. GV was involved in providing the idea for this paper, producing the COLA and N-body MG simulations and contributed to the writing of the text. MI was involved in developing the idea for the paper, the MG part of the design, contributed to the analysis of Fisher and MCMC results, mentored the project, and contributed to the text of the paper. KH was involved in developing the idea for the paper, provided the design, and contributed to the text of paper.

LSST DESC acknowledges ongoing support from the Institut National de Physique Nucl\'eaire et de Physique des Particules in France; the Science \& Technology Facilities Council in the United Kingdom; and the Department of Energy, the National Science Foundation, and the LSST Corporation in the United States. LSST DESC uses the resources of the IN2P3 / CNRS Computing Center (CC-IN2P3--Lyon/Villeurbanne - France) funded by the Centre National de la Recherche Scientifique; the Univ. Savoie Mont Blanc - CNRS/IN2P3 MUST computing center; the National Energy Research Scientific Computing Center, a DOE Office of Science User Facility supported by the Office of Science of the U.S.\ Department of Energy under contract No.\ DE-AC02-05CH11231; STFC DiRAC HPC Facilities, funded by UK BIS National E-infrastructure capital grants; and the UK particle physics grid, supported by the GridPP Collaboration.  This work was performed in part under DOE contract DE-AC02-76SF00515.

Argonne National Laboratory's work was supported under the
U.S. Department of Energy contract DE-AC02-06CH11357. 
GV recognizes financial support by DoE grant DE-SC0011838, NASA ATP grants
NNX14AH53G and 80NSSC18K0695, NASA ROSES grant 12-EUCLID12-0004 and funding related to the WFIRST Science Investigation Team.
MI acknowledges that this material is based upon work supported in part by the Department of Energy, Office of Science, under Award Number DE-SC0019206.

We would like to thank Salman Habib and Eske Pedersen for useful discussions related to this project. We also would like to thank Baojiu Li, for kindly providing us with the MG N-body simulation data. GV would also like to thank Rachel Bean and Hans Winther for useful discussions related to this project. NR would like to thank Andrew Hearin for assistance with Python packaging.

The emulator is built using the following Python packages: \verb|GPflow| \cite{GPflow2017}, \verb|TensorFlow| \cite{tensorflow} and \verb|Scikit-learn| \cite{scikit}. The analyses performed in this paper utilize the following:  \verb|Numpy| and \verb|Scipy| \cite{numpyscipy}, \verb|Matplotlib| \cite{matplotlib}, \verb|emcee| \cite{emcee} and \verb|GetDist| \cite{getdist}. The final trained emulator is provided at \url{https://github.com/LSSTDESC/mgemu} \cite{valogiannis_georgios_2020_4058880}.

\input{ms.bbl}
\end{document}

%% file: introduction_sec.tex
The Vera C. Rubin Observatory Legacy Survey of Space and Time (LSST) \footnote{\url{http://www.lsst.org}} \citep{Abell:2009aa,Abate:2012za}, together with a wide range of current and future surveys of the large-scale structure (LSS) of the universe, such as DESI \citep{Levi:2013gra}, Euclid \citep{Laureijs:2011gra}, the Nancy Grace Roman Space Telescope \citep{Spergel:2013tha} and SPHEREx \citep{Dore:2014cca}, will offer a unique opportunity to test our standard cosmological assumptions at an unprecedented level of accuracy. The widely accepted cosmological model, $\Lambda$ Cold Dark Matter ($\Lambda$CDM), attributes the observed accelerated expansion of the universe \cite{Perlmutter:1998np, Riess:2004nr} to the existence of a positive cosmological constant, $\Lambda$, corresponding to non-zero vacuum energy. This assumption, combined with the existence of pressure-less cold dark matter and gravity described by Einstein's General Relativity (GR), has been very successful at fitting a large spectrum of cosmological observations \citep{Eisenstein:2005su, Percival:2007yw, Percival:2009xn, Kazin:2014qga, Ade:2013zuv, Ade:2015xua}.  

Despite these remarkable observational accomplishments, $\Lambda$CDM has faced several theoretical challenges, with the cosmological constant problem \citep{Weinberg:1988cp,Carroll2001} arguably serving as the primary reason to consider alternative proposals. Furthermore, as our ability to accurately obtain the underlying cosmological parameters from late and early-time observations increases, attention has been drawn to certain tensions between the corresponding extracted values of the Hubble constant, $H_0$ \citep{10.1093/mnras/stu278,Riess_2019,10.1093/mnras/stx1600,LOMBRISER2020135303}, and the amplitude of density fluctuations, $\sigma_8$ \citep{10.1093/mnras/stt601,10.1093/mnras/stw2805,Abbott:2017wau,Ade:2015xua,PhysRevD.96.083532}, which could however be attributed to unknown systematics. 
Combined with the long-term interest in exploring deviations from GR in all regimes \citep{Will:2005va,Will:2014kxa}, the above motivate introducing large-scale modifications of gravity as alternative candidates for cosmic acceleration; such theories are called the Modified Gravity (MG) theories \citep{Koyama:2015vza,Clifton:2011jh, Ishak:2018his,Ferreira:2019xrr}.

In order to be able to evade the existing tight constraints of gravity from observations in the Solar System \citep{Will:2005va,Will:2014kxa}, while at the same time producing detectable large-scale signatures, viable MG candidates typically invoke ``screening'' mechanisms \citep{Khoury:2010xi,Khoury:2013tda}, which suppress deviations in the high-density regime through novel scalar field self-interactions \citep{VAINSHTEIN1972393,Babichev:2013usa,PhysRevD.69.044026,PhysRevLett.93.171104,PhysRevLett.104.231301,Olive:2007aj,Dvali:2010jz}. Furthermore, the space of all MG parametrizations that lead to second order equations of motion, the Horndeski class \citep{Horndeski1974,PhysRevD.84.064039,Kobayashi:2011nu} has been additionally restricted \citep{Lombriser:2015sxa,Lombriser:2016yzn,Sakstein:2017xjx,Ezquiaga:2017ekz,Creminelli:2017sry,Baker:2017hug} by the simultaneous detection of gravitational waves and electromagnetic counterparts by the LIGO/Virgo collaborations \citep{TheLIGOScientific:2017qsa,Goldstein:2017mmi,Savchenko:2017ffs,Monitor:2017mdv,GBM:2017lvd}. A detailed discussion of viable MG candidates testable by LSST Dark Energy Science Collaboration (DESC) was presented in Ref.~\cite{Ishak:2019aay}.

The predicted transition from MG to GR would manifest itself, by means of the dynamical screening mechanism, in the nonlinear regime of structure formation, which will be precisely probed by the Stage-IV surveys of the LSS \citep{Albrecht:2006um}. As a result, these upcoming observations will offer a unique opportunity to study the large-scale behavior of gravity with unprecedented accuracy. The optimal interpretation of the wealth of upcoming data, however, is conditional upon our ability to produce efficient and reliable theoretical predictions of the expected observable signatures of MG. In the (quasi-)linear regime of structure formation, this can be partially achieved through analytical, perturbation theory approaches, such as Lagrangian Perturbation Theory (LPT) \citep{Aviles:2017aor,Aviles:2018saf,Aviles:2018qot,Valogiannis:2019xed,Valogiannis:2019nfz}. Unfortunately, these approaches break down on nonlinear scales. Reliable predictions of such small-scale signals from MG theories can only be obtained through performing full N-body simulations (for a comparison of different codes, see Ref.~\citep{Winther:2015wla}). These are computationally very  expensive, particularly in the presence of an additional MG-induced force, due to the inherent nonlinearities introduced by the screening mechanism.

As a consequence of the substantial computational costs associated with performing full N-body simulations for MG models, efficient approaches are essential. To that end, the hybrid COmoving Lagrangian Acceleration (COLA) method was developed in Ref.~\citep{Valogiannis:2016ane}, expanding upon the initial $\Lambda$CDM implementation of Ref.~\citep{Tassev:2013pn}, for efficiently simulating the MG classes of chameleon and symmetron screening.  Utilizing a combination of $2^{\rm nd}$ order Lagrangian Perturbation Theory (2LPT) and a pure N-body component, the COLA method provides a great trade-off between accuracy and computational efficiency, that was found to recover the fractional deviation in the nonlinear dark matter power spectra with sufficient accuracy, at only a fraction of the standard computational cost.
However, many investigations require even faster prediction capabilities to enable the exploration of parameter space. For example, Markov Chain Monte-Carlo (MCMC) inference relies on tens of thousands of model evaluations and even the fast COLA approach, which takes $\sim 1$ hour on a single core for one MG simulation, would be too slow to enable such an investigation.

To provide faster predictions for, e.g., the power spectrum, fitting functions have been used extensively in the past (see, e.g., Refs.  \cite{2003MNRAS.341.1311S,2012ApJ...761..152T} for fits capturing $\Lambda$CDM and $w$CDM cosmologies based on the Halofit approach  or a halo model-based approach to include baryonic effects~\cite{Mead_2015} or physics beyond $\Lambda$CDM~\cite{Mead_2016}). However, fitting functions have several drawbacks. First, the accuracy requirements for ongoing and future surveys of a few to sub-percent are very difficult to obtain by a single functional form and a set of fitting parameters. Second, for models outside the range for which the fit was derived, no error bounds are available and biases can occur. Third, a large range of simulations is needed to enable a good calibration of the parameters that describe the fit. Overall, it has been shown that even for the relatively restricted case of $w$CDM cosmologies, it is very difficult to achieve an accuracy of better than 5--10\% as pointed out in one of the recent Halofit papers~\cite{2012ApJ...761..152T} and subsequent comparisons, e.g., Refs.~\cite{2017ApJ...847...50L,euclidemu2019}.

Because of the above shortcomings of fitting functions and the need for very accurate and fast predictions, the concept of emulators was introduced to cosmology in Refs.~\citep{2006ApJ...646L...1H,2007PhRvD..76h3503H}. It was shown that with a relatively small number of high-quality simulations, prediction schemes could be built that provide high-accuracy results for, e.g., the matter power spectrum and $C_\ell$s quickly. The \verb|Coyote Universe| project~\citep{2010ApJ...715..104H,2009ApJ...705..156H,2010ApJ...713.1322L} then released a stand-alone prediction scheme for the matter power spectrum, the \verb|CosmicEmu|, based on a set of highly accurate simulations. The work was extended in several ways, including the coverage of a larger $k$-range and redshift and parameter spaces~\citep{2014ApJ...780..111H,2017ApJ...847...50L}. The emulation concept itself has since then become rather popular and was used in several studies concerning the matter power spectrum, see, e.g. Refs.~\cite{Agarwal_2014,euclidemu2019}, a range of other summary statistics, e.g., galaxy power spectra~\cite{Kwan_2013,Wibking_2019}, concentration mass relation~\cite{Kwan_2013}, the halo mass function~\cite{bocquet2020miratitan} as well as comprehensive simulation efforts that extracted a range of emulators, such as the {\verb|AEMULUS| project}~\cite{DeRose_2019,McClintock_2019,Zhai_2019,mcclintock2019aemulus} and {\verb|DARK QUEST|}~\cite{Nishimichi_2019,kobayashi2020accurate}.

In this work we develop a Gaussian Process emulator to estimate the fractional deviation of the nonlinear matter power spectrum $P_{MG}(k)/P_{\Lambda CDM}(k)$ (also referred to as enhancement/boost in the power spectrum or the power spectrum ratio) as a function of cosmological parameters and redshift, trained on a set of COLA simulations in the MG scenario. Given that our focus is to efficiently emulate statistics for MG models that will be testable by LSST DESC, our target model needs to be one that exhibits a well-studied phenomenology (including existing full N-body simulations) and will predict detectable deviations in the scales of interest to the survey. For this reason our chosen candidate is the $f(R)$ Hu-Sawicki model \cite{Hu:2007nk}, which realizes the chameleon screening mechanism and which we have previously identified as one of the prioritized beyond-$w(z)$CDM candidates testable by LSST DESC \cite{Ishak:2019aay}. 

In this paper we discuss the construction of the emulator that comprises of an experimental design, training data synthesis and the statistical techniques to perform interpolation across cosmological parameters and redshifts. After training the emulator on the COLA-generated dataset, we validate its ability to recover simulated test cosmologies within our target range, before proceeding to compare its accuracy against results obtained by full N-body simulations for the Hu-Sawicki model. Having quantified its accuracy, we then illustrate the capabilities of our emulator, which delivers a massive speed-up by 6 orders of magnitude over COLA simulations, through three essential applications: a sensitivity analysis, obtaining parameter constraints through Fisher forecasting and MCMC inference. It is worth noting, at this point, that efficient predictions for power spectra in the Hu-Sawicki MG model have also been recently presented in Refs.~\cite{Winther2019, Giblin2019,Bose:2020wch}. However, our work is clearly different in that, while these works presented predictions through appropriately designed fitting formulas \cite{Winther2019} or semi-analytical models \cite{Giblin2019,Bose:2020wch}, our emulator is trained directly from simulations, expanding upon the established $w$CDM infrastructure of the \verb|CosmicEmu| \citep{Heitmann:2013bra}. 

Our paper is structured as follows: in Sec.~\ref{Training}, 
we describe the approach to generating our training data set by first introducing our target MG model, then discussing our choices for the experimental design, including cosmological parameters and their ranges, and finally describing the efficient COLA approach we use to generate the simulations.
Next, we discuss in Sec.~\ref{Development} the details of our emulator development. We then proceed in Sec.~\ref{results_sec} to validate the emulator results, before presenting three  applications in Sec.~\ref{applications_sec}. Finally, we conclude and discuss future work in Sec. \ref{conclusions_sec}. Technical details on Gaussian Processes and the emulator design can be found in Appendices \ref{Gaussian_app}  and \ref{sim_app}, respectively.

%% file: data_sec.tex
\subsection{Modified Gravity Model}\label{MGmodel}

Theoretical investigations of potential departures from Einstein's GR, as well as of their consequent observational implications \citep{Will:2005va}, have been an active research topic, particularly in the last two decades, due to their potential implications on resolving the mystery of cosmic acceleration \citep{Koyama:2015vza, Ishak:2018his,Ferreira:2019xrr}. One of the most commonly considered classes of MG deviates from GR through the addition of a nonlinear function $f(R)$ of the Ricci scalar $R$ to the standard Einstein-Hilbert action. These are the $f(R)$ gravity theories \citep{DeFelice:2010aj,Burrage2018}, which are described by the following action $S$:
\begin{equation}\label{actFr}
S=\int d^4x \sqrt{-g} \left[\frac{R+f(R)}{16 \pi G} + \mathcal{L}_m \right],
\end{equation}
with $\mathcal{L}_m$ denoting the matter sector Lagrangian, $G$ the gravitational constant and using units where the speed of light in vacuum, $c$, has been set to unity. The modification of the form (\ref{actFr}) activates, in principle, a new additional degree of freedom in the gravitational sector, which could be responsible for driving the cosmic expansion to accelerate, instead of dark energy \citep{Carroll:2003wy}. 

The most widely-considered model of this type is the $f(R)$ Hu-Sawicki model \citep{Hu:2007nk}, with the functional form:
\begin{equation}\label{fRHu}
f(R)=-m^2\frac{c_1\left(R/m^2\right)^n}{c_2\left(R/m^2\right)^n+1},
\end{equation}
where in eq. (\ref{fRHu}) $m=H_0\sqrt{\Omega_{m}}$, with $\Omega_{m}$ the matter fractional energy density and $H_0$ the Hubble Constant, both evaluated today and $n$, $c_1$ and $c_2$ are free parameters of the model. 

Any well-motivated MG parametrization should have the flexibility to match the observed expansion history well-described by $\Lambda$CDM, a requirement which gives (for sufficiently small values of $|f_{R_0}|$), the following relationship for the background value of the Ricci scalar, $\bar{R}$:
\begin{align}\label{constr1}
\bar{R}= 3 \Omega_m H_0^2 \left(1+ 4 \frac{\Omega_{\Lambda}}{\Omega_{m}} \right),
\end{align}
and for the derivative $f_{R}=\frac{df(R)}{dR}$ 
\begin{equation}\label{scalaron}
\bar{f}_{R_0} =-n\frac{c_1}{c_2^2}\left(\frac{\Omega_{m}}{3(\Omega_{m}+\Omega_{\Lambda})}\right)^{n+1},
\end{equation}
where $\Omega_\Lambda$ is the dark energy fractional density evaluated at the present time. 
Through relationship (\ref{scalaron}), one can reduce the number of free parameters of the Hu-Sawicki model, which can be fully characterized by the pair \{$|f_{R_0}|,n$\}. We briefly point out here that in the limit of $|f_{R_0}|\rightarrow 0$ the background cosmology of $\Lambda$CDM is recovered. Through an appropriate conformal transformation, the model can be cast into a scalar-tensor theory (with $f_{R}$ acting as the MG scalar field) \citep{Brax:2008hh} that recovers GR in regions of large Newtonian potential through the chameleon screening mechanism \citep{PhysRevD.69.044026,PhysRevLett.93.171104}.

The Hu-Sawicki model produces novel, distinct signatures which are testable by upcoming cosmological surveys and as a result has been well-explored in the literature through full N-body simulations. It is also known to be free from any physical instabilities \cite{Clifton:2011jh}. We further note that, despite the increasingly tight observational constraints placed on it over the past decade (see, e.g., \citep{Burrage2018} for a review), the Hu-Sawicki model remains viable. For all these reasons, 
it serves as an ideal test-bed for investigating cosmological theories of gravity and is one of the main candidate models to be considered by DESC \citep{Ishak:2019aay}. The referenced document describes the various beyond-w(z) CDM model prioritized for study by DESC. 

\subsection{Experimental Design}

Two main criteria govern the choice of the cosmological parameter space covered by our emulator -- first, a consideration of the computational cost incurred in generating a set of COLA simulations, which will determine how many models can be reasonably run, and, second, an estimate for the emulator's target accuracy. The application of these two criteria determines how many parameters we can afford to include and how broad their priors can be. 

Our main interest in this paper is the exploration of the two parameters that define the Hu-Sawicki model for $f(R)$ gravity theories, \{$f_{R_0},n$\}. We aim to vary these parameters over a wide range in order to span a broad array of deviations predicted by the model. We choose  
\begin{eqnarray}\label{equ:params1}
		-8 \leq & \log(f_{R_0}) &\leq -4, \\ 
		0 \leq & n &\leq 4
\end{eqnarray}

and note that the upper end of the chosen $f_{R_0}$ range corresponds to a large modification case in which screening is absent, whereas for the low end value, modified gravity forces are very strongly suppressed \citep{Hu:2007nk,Zhao:2010qy}. Most studies in the MG literature, e.g., Refs.~\cite{Winther2019,Bose:2020wch}, restrict their attention on the subspace of fixed $n=1$ and only vary $f_{R0}$, given that deviations are more sensitive to the latter parameter. In this work, however, we also consider variations with respect to $n$, thus allowing a more complete theoretical exploration of the Hu-Sawicki model parameter space.  
Given the wide dynamical range of $f_{R_0}$, over-sampling of large parameter values may occur if the parameter range is sampled linearly. In order to mitigate this problem, we instead use a logarithmic scaling in the experimental design. 

Next, we have to decide which additional parameters we want to vary for the CDM component of the model. We want to focus on the parameters that affect the power spectrum the most and at the same time aim to vary only a small number of parameters. An increase in the number of cosmological parameters varied leads to either a less accurate emulator or a much larger number of simulations needed. It is therefore desirable to keep the number of parameters varied small while keeping the needed accuracy.

Figure~10 in Ref.~\cite{2009ApJ...705..156H} shows a sensitivity analysis for the five parameters of the $w$CDM ``Standard Model'' of cosmology. The image shows the variation of the ratio of the power spectrum to a model that is evaluated at the midpoint of the five Standard Model parameters changing one parameter at a time. The baryon density $\Omega_b$ clearly does not have much of an effect on the power spectrum ratio, while \{$\Omega_m h^2, \sigma_8,n_s$\} all show considerable impact on particularly large scales. While the dark energy equation of state also leads to considerable variations, $w$ is not of interest here as we fix the background of a $\Lambda$CDM model. We note again that the requirement to match a $\Lambda$CDM background expansion imposes additional restrictions on the parameter space of the model, which are reflected in Equations (\ref{constr1}) and (\ref{scalaron}) that we introduced Sec~\ref{MGmodel}. As we briefly discuss at the end of Section~\ref{sec:nbodycomp}, variations of $h$ given current observational bounds do not affect the fractional deviation of the nonlinear matter power spectrum $P_{MG}(k)/P_{\Lambda CDM}(k)$ considerably. We therefore allow variations of the aforementioned additional three parameters for our emulator. In future work, the parameter space will be extended.

We now proceed to set the range for three of the standard CDM parameters \{$\Omega_m h^2, \sigma_8,n_s$\}. We choose the same range used as in the Coyote Universe~\cite{2009ApJ...705..156H} and the Mira-Titan Universe simulation suites \cite{2016ApJ...820..108H}. Both papers provide in-depth discussions about the choices to balance broad parameter coverage and achievable accuracy goals for the emulator, as informed by contemporary and future cosmic microwave background (CMB) and LSS surveys. Choosing the same parameter ranges also has the advantage that results in the future can be compared to the previous work and possibly combined later on. In summary, for the three parameters, we choose:
\begin{eqnarray}
		0.12 \leq & \Omega_m h^2 &\leq  0.15, \\ 
		0.85 \leq & n_s & \leq  1.1, \\ 
		0.7 \leq  &  \sigma_8 &\leq  0.9 \label{equ:params2}
\end{eqnarray}
We add that $\Omega_b h^2 = 0.0223$ (while we ignore the effects of massive neutrinos) and that for the dimensionless Hubble constant we have chosen $h=0.67$. After having decided on the five parameters and their ranges we have to determine the number of simulations needed and pick a sampling scheme that allows us to set up a suitable parameter sampling design. Past experience has shown that roughly 10 simulations per parameter are needed to enable the construction of an accurate emulator (targeting few percent accuracy), leading to a set of $\sim 50$ COLA simulations.

The next choice to be made concerns the employed sampling scheme. For an excellent discussion on different sampling schemes used in computer experiments, the interested reader is referred to Ref.~\cite{Santner03}. Ref.~\cite{2009ApJ...705..156H} provides an extensive discussion about different methods in the cosmology context.  In this paper, we use a symmetric Latin hypercube (SLH) design, introduced in Ref.~\cite{YE2000145}. Latin hypercube (LH) sampling schemes are statistical stratified sampling methods used to generate near-random samples of values from a multi-dimensional distribution, such that there exists only one sample in each sub-division for each parameter range. Compared to a random space-filling scheme, which does not take into account the previous sampled points in a new sample-point generation, an LH sampling strategy guarantees an optimal representation of the variability of parameters. While sampling on a uniform grid also ensures fair representation, the number of required simulations would be prohibitively large. The SLH offers a space-filling design strategy that allows for flexibility with regard to a number of design points and is computationally inexpensive when optimizing the design itself, when compared to other methods such as Orthogonal Array LH implementations. The SLH imposes additional, specific symmetry requirements compared to other LH designs, as described in detail in e.g.~Ref.~\cite{2009ApJ...705..156H}. Ref.~\cite{2009ApJ...705..156H} provides concrete examples to  illustrate these symmetries. The symmetry imposes a space-filling requirement on the designs considered upfront, which  carries through to  all  projections. 

The final choice concerns the proposal distribution for the cosmological parameters. We use uniform distributions for the cosmological parameters $\theta = \{\Omega_m h^2, \sigma_8, n_s, \log(f_{R_0}),n\}$ to ensure an unbiased exploration of parameter space. The 50 cosmological models that are chosen using the above prescription are shown in Appendix \ref{sec:params_appendix}. In other efforts, the design has been created based on posteriors from surveys see, e.g., the \verb|AEMULUS| project~\cite{2019ApJ...875...69D}. This approach reduces the number of required simulations but also restricts the viability of the emulator considerably, because of the limited effective sampling volume and because of potential biases introduced by weighted sampling.   

\subsection{COLA Simulations with Modified Gravity}\label{sec:COLAsims}

\begin{figure}
	\includegraphics[width=\columnwidth]{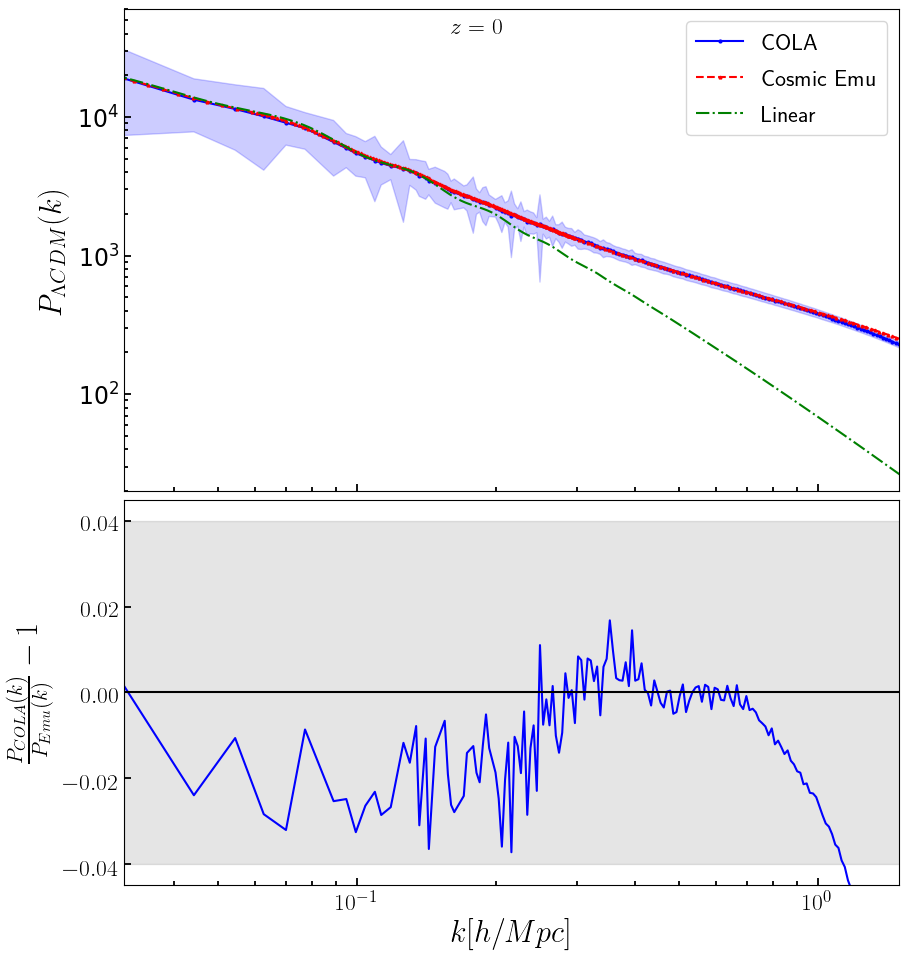}
    \caption{Top: $\Lambda$CDM matter power spectra for the fiducial cosmological parameters at $z=0$, as obtained by the efficient COLA method (blue solid line) and \texttt{ CosmicEmu} [red dashed line]. The linear matter power spectrum for the same cosmology [green dot-dashed line] is also shown. The shaded blue region represents the standard deviation obtained from the 2000 available $P(k)$ COLA realizations. Bottom: Fractional difference between the COLA and the \texttt{CosmicEmu}-generated power spectra of the upper panel. The shaded gray region highlights the $4 \%$ level of accuracy.}
    \label{fig:COLA_validation}
\end{figure}

The signatures introduced by MG models manifest themselves in the nonlinear regime of structure formation, where the screening mechanism is in full effect, as well as in the intermediate quasi-linear scales. As a result, perturbation theory approaches fail to capture the full picture of structure formation in the presence of a modification to gravity, which can only be performed through N-body simulations. These are particularly computationally expensive, due to the inherent non-linearity of the scalar-field Klein-Gordon equation. An overview and comparison of different full N-body codes in MG can be found in Refs.~\citep{Winther:2015wla}.

Given the substantial computational cost of running multiple full N-body simulations to train our emulator, we instead utilize the Comoving Lagrangian Acceleration (COLA) hybridization scheme for efficient simulations of chameleon and symmetron screening models developed in Ref.~\citep{Valogiannis:2016ane}, expanding upon the initial $\Lambda$CDM implementation of Ref.~\citep{Tassev:2013pn}. Through a combination of $2^{\rm nd}$ order Lagrangian PT (2LPT) that evolves the large scales and a pure N-body component for integrating the nonlinear regime, the COLA approach was found to be able to recover the nonlinear dark matter power spectrum (in $\Lambda$CDM) using only a few tens of timesteps \citep{Tassev:2013pn}. The implementation developed for the $f(R)$ Hu-Sawicki model in Ref.~\citep{Valogiannis:2016ane}, which is the one we employ here, was shown to successfully model the fractional deviation in the dark matter power spectrum, $P_{MG}(k)/P_{\Lambda CDM}(k)$, in the nonlinear regime, using an effective ``thin-shell'' approach for capturing the chameleon screening effect that was presented in  Ref.~\citep{Winther:2014cia}. The reported accuracy in the estimation of this power spectrum ratio was at the level of $1$ percent, compared against N-body simulations at $z=0$. Below we summarize the parameters we choose for our COLA runs, while more details about the COLA MG implementation can be found in Ref.~\citep{Valogiannis:2016ane}.

\begin{table}
\caption{Parameters of the COLA MG simulation suite.}
\centering
{
\begin{tabular}{lc}
\hline
Box size, $L$ & 200 $h^{-1}$Mpc \\
Number of particles, $N_p$ & $256^3$\\
Number of grids, $N_g$ & $512^3$\\
Initial redshift $z_i$ & $49$ \\
Final redshift $z_f$ & $0$ \\
Number of time-steps, $N_t$ & $100$\\
Number of realizations, $N_r$ & $1$\\
Dimensionless Hubble constant, $h$ & $0.67$\\
\hline
\end{tabular}%
}
\label{T_sims}
\end{table}

Our simulations are initialized using the 2LPT initial conditions code (\verb|2LPTic|) \citep{Scoccimarro:1997gr} at an initial redshift of $z_i=49$. Given that the effects of MG are assumed to be negligible at early times within the context of cosmic acceleration, we use the same set of $\Lambda$CDM initial conditions for both the $\Lambda$CDM and the $f(R)$ runs, as in Ref.~\citep{Valogiannis:2016ane}. For each of the (50+5) cases in our experimental design (listed in Table \ref{T_Mill}), the linear $\Lambda$CDM matter power spectrum is produced with \verb|CAMB|~\citep{Lewis:1999bs}, which is used to generate the initial conditions with \verb|2LPTic| at $z_i=49$. The COLA simulations are then run using the parameters of Table \ref{T_sims}, for both the $\Lambda$CDM and the $f(R)$ Hu-Sawicki case with the same initial random seed. At each of the 100 timesteps of the simulations, the matter power spectra are stored, for both $\Lambda$CDM and the MG case, with $213$ bins equally spaced logarithmically in the $k$-range of $(0.03--3.5) h$Mpc$^{-1}$. The choice of 100 timesteps is made so that the target redshift range of our emulator is adequately spanned using only a single run for each case, without at the same time making the simulations too time-consuming \footnote{We should note here that if one wants to make predictions for an individual redshift, the efficient COLA approach was shown to work well with much fewer timesteps \citep{Tassev:2013pn,Valogiannis:2016ane}}. Furthermore, the high number of $k$ bins ($n_{bins} = 213)$ is chosen to guarantee the accuracy of the redshift interpolation. 
Finally, the exact same specifications are used in the $2000$ simulations we perform, each of them for a different randomly chosen initial seed, in order to obtain the covariance matrix of the ratio for the fiducial cosmology of Section \ref{Fishersubsection}. 

Before we discuss COLA's accuracy in predicting summary statistics in MG (the matter power spectrum ratio here), which will be addressed in detail in Sec.~\ref{sec:nbodycomp}, we start by presenting the $\Lambda$CDM benchmark in Fig.~\ref{fig:COLA_validation}. We find that the mean of the 2000 COLA-generated $\Lambda$CDM power spectra for the fiducial cosmology remains consistent, within $\sim 3\%$, with the nonlinear \verb|CosmicEmu| prediction for the same cosmology down to $k\sim 1 h$Mpc$^{-1}$. It is worth noting that the better agreement compared to the initial COLA implementation in Ref.~\citep{Tassev:2013pn}, is attributed to the fact that we are using about 3 times as many time-steps as the runs in that original work, for the reasons discussed in the previous paragraph. 

Given that the COLA method is known to perform better at recovering the fractional deviation $P_{MG}(k)/P_{\Lambda CDM}(k)$, rather than the power spectrum itself \citep{Valogiannis:2016ane}, and also because this quantity is much less sensitive to cosmic variance effects, we choose these ratios as our training data in the Gaussian process emulator. Modeling this quantity has also been the target of interest by other studies in the MG literature \citep{10.1093/mnras/stz1836,PhysRevD.100.123540,Bose:2020wch}. The residual noise in the ratio is later smoothed out using a Savitzky-Golay filter, as explained in Sec.~\ref{sec:development}.

%% file: development_sec.tex
\label{sec:development}

Based on a carefully chosen design strategy to determine a set of training points, a well-matching interpolating strategy can be selected in order to estimate the summary statistics at intermediate cosmologies. Neural networks \citep{Agarwal_2014}, polynomial chaos expansions \cite{euclidemu2019} and Gaussian processes \cite{2006ApJ...646L...1H,2007PhRvD..76h3503H, 2010ApJ...715..104H,2009ApJ...705..156H,2010ApJ...713.1322L} have been successfully employed to construct emulators for the prediction of astrophysical summary statistics. In particular, Gaussian Processes are an attractive way of performing machine learning tasks with small number well sampled data points. This non-parametric Bayesian regression method provides fast, interpretable and high-fidelity estimations with associated uncertainties. For these reasons, we utilize Gaussian processes, along with Principal Component Analysis and linear interpolation schemes to construct our emulator. 

\subsection{Gaussian Process Interpolation across Cosmological Parameters}

Our emulation strategy for the cosmological parameters $\theta$  follows a similar routine employed for the \verb|CosmicEmu| \cite{Heitmann2006} construction, using Gaussian processes (GPs) in a representation space \cite{Higdon2008}. The individual steps, including the data pre-processing are as follows:

\begin{enumerate}

\item The individual ratios of the power spectra  are noisy since we only perform one realization for each individual COLA simulation. Emulators designed directly based on this data may pick up undesired noise from the data. To avoid this problem, we utilize the Savitzky-Golay filter (or savgol filter, \cite{Savitzky1964, Press1990}) to obtain a smoothed power spectrum ratio $\chi(k)$, as detailed in Appendix \ref{sec:smooth_appendix}.

\item A standardization transformation on both smoothed power spectrum ratio and cosmological parameters is performed, to result in a mean of zero and a standard deviation of one for their respective distributions: 
\begin{eqnarray}
    \theta_i{'} &=& (\theta_i - \mu_{\theta_i})/\sigma_{\theta_i}, \\
    \chi{'}(k) &=& (\chi(k) - \mu_{\chi(k)})/\sigma_{\chi(k)}.
    \label{eq:scale}
\end{eqnarray}
The standardized power spectrum ratio $\chi{'}(k)$ is re-scaled using the mean $\mu_{\chi(k)}$ and standard deviation $\sigma_{\chi(k)}$. The mean and standard deviations are computed collectively for all the 50 cosmological models. Similarly the individual cosmological parameters $\theta_i$ are re-scaled to $\theta_i{'}$ by their means $\mu_{\theta_i}$ and standard deviations  $\sigma_{\theta_i}$.   

\item A Singular Value Decomposition (SVD) is performed on the smoothed and normalized power spectrum enhancement $\chi(k, \theta)$ for dimensionality reduction. This is a generalization of eigenvalue decomposition to any rectangular matrix, whereby a matrix is factorized into a set of orthonormal vectors.  Equation \eqref{eq:pca} below shows the decomposition to  the basis $\phi_m(k)$ and weights $w_m(\theta)$ of the representation, truncated at $n_w$ eigenvectors:  

\begin{equation}
    \chi{'}(k, \theta{'}) = \sum_{m=1}^{n_w} \phi_m(k) w_m(\theta{'}) + \epsilon.
    \label{eq:pca}
\end{equation}

The excess information that is not captured by this decomposition is represented by $\epsilon$.
The Principal Component Analysis (PCA) of the power spectrum enhancement reveals that a total of $n_w = 6$ eigenvectors successfully capture over $99.99$ percent of the variance in the data, effectively allowing us to truncate the expansion without significant loss of information. In addition to dealing with a reduced dimension (from $n_{bins}=213$ to $n_w=6$) of eigenvectors, this also enables orthogonality in the interpolation space, i.e., the new basis $\phi_m(k)$ that maximizes variance is an uncorrelated representation of $\chi{'}(k, \theta)$.

\item The weights $w_m(\theta)$ corresponding to $n_w = 6$ truncated orthogonal bases $\phi_m(k)$ are then modeled as functions of the input parameters $\theta$. This local interpolation in parameter space is made using multivariate Gaussian Process regression applied to the weights of the Principal Component bases, as explained in Appendix \ref{sec:gp_appendix}.

Configuration  of  the  covariance function and determination of the associated hyperparameters are the key components for learning the correct GP fit. We choose a popular Matern-5/2 kernel \cite{Rasmussen2006}, and check for robustness of the emulator accuracy with different choices of covariance functions. We search for the best combination of hyper-parameters of the GP using a fast gradient-based optimization called Adagrad \cite{Duchi2011}. 

\end{enumerate}

The four steps above are applied to all the 100 snapshots, resulting in a suite of 100 emulators for smoothed power spectrum deviations. For any new cosmological parameters $\overline{\theta}$, the trained GP generated 6 corresponding weights, and these are multiplied by the PCA basis vectors to generate new power spectrum enhancement values with the $k$-range of the emulator. 

\subsection{Redshift Interpolation}

The sampling for the redshifts coverage is treated separately from the sampling of the cosmological parameters. 
The cosmological parameter values are generated using an SLH design to ensure representation across the five dimensional parameter space. 
In contrast, the COLA snapshots are all created at the same redshifts between $z_i = 49$ and $z_f = 0$ with linear spacing in corresponding scale factors. Due to this difference in the training sampling, we do not employ the same interpolation scheme for all the parameters. Instead a separate interpolation routine is executed between the outputs of independent GP emulators. Equation \eqref{eq:linear} shows a simple linear interpolation for an intermediate redshift $z$ when emulator results at two nearest redshifts $z_{-}$ and $z_{+}$ are calculated from the previous section: 
\begin{multline}
\label{eq:linear}
    \chi_{emu}(k; \theta, z) = \chi(k; \theta, z_{-}) \\
    + \frac{z - z_{-}}{ z_{+} - z_{-}}
    (\chi(k; \theta, z_{+}) - \chi(k; \theta, z_{-})).
\end{multline}
With our emulator suite for 100 individual redshifts, a simple linear interpolation works within one percent  accuracy for the redshift interpolation when the closest two emulators are used for a given cosmology. Although this requires two independent GP evaluations and PCA reconstructions, this final emulator $\chi_{emu}(k, \Omega_m, \sigma_8, n_s, \log(f_{R_0}),n, z)$ is found to be robust and fast.

%% file: results_sec.tex
The accuracy of our emulator is determined by two factors. First, limitations of the underlying simulations lead to irreducible errors. We have chosen to use Lagrangian Perturbation Theory for the cosmological simulations over a computationally expensive full N-body alternative. This choice restricts the accuracy of the emulator on small scales. Second, an error arises due to the limited number of training samples and the nature of the sampling and interpolation schemes.  

We study these effects by carrying out two types of verifications. For the first test we compare the emulator against a set of six additional COLA simulations that are not part of the design. This allows us to evaluate the accuracy of the emulator construction itself. 
The second test utilizes three state-of-the-art N-body simulations for beyond $\omega(z)$CDM cosmologies and we compare the emulator performance directly to the measurements from the simulations. This test allows us to evaluate the overall performance of the emulator, including both errors from the limited simulation accuracy and the emulator construction itself.

We restrict the emulator predictions to $k\le 1 h$Mpc$^{-1}$. This choice is mainly driven by the accuracy of the COLA simulations. As shown in Fig.~\ref{fig:COLA_validation}, the COLA approach is in very good agreement with measurements from high-resolution N-body simulations, represented by the \verb| CosmicEmu| result, out to $k\sim 1 h$Mpc$^{-1}$. At this point, the COLA power spectrum starts to deviate from the \verb| CosmicEmu| result and underpredicts the power spectrum at the few percent level. Restricting our emulator out to this $k$-range therefore seems well justified. In addition, beyond $k\sim 1 h$Mpc$^{-1}$ other effects, like baryonic physics become more and more important (see, e.g., Ref.~\cite{2018MNRAS.475..676S} for a recent discussion of the effects of baryonic physics on the power spectrum).

\subsection{Comparison with COLA Simulations}\label{sec:holdout}

\begin{figure}
	\includegraphics[width=\columnwidth]{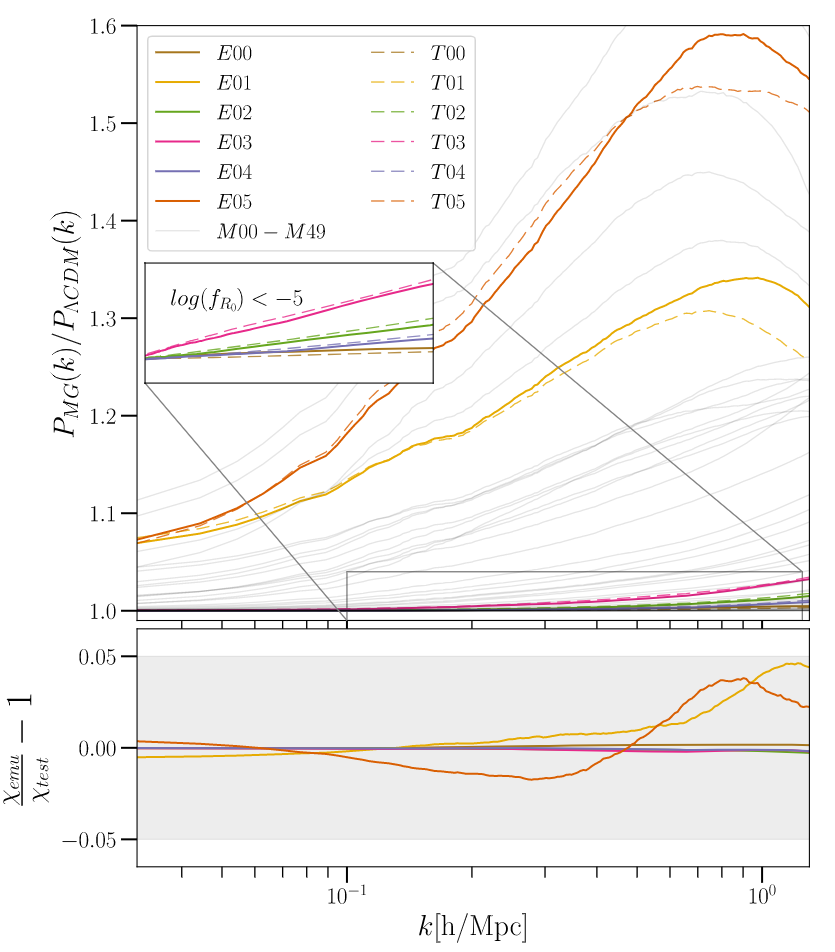}
    \caption{Test of the emulator accuracy for additional cosmologies. Top: Power spectrum ratios $P_{MG}(k)/P_{\Lambda CDM}(k)$ for the six testing COLA simulations at $z=0$, $T00-T05$, are compared to the corresponding emulator results, $E00-E05$.  Training cosmologies, $M00-M49$, are shown in gray. Bottom: Relative error of the emulator prediction compared to the COLA simulations. For the models with large values for $f_{R_0}$ ($T01$ and $T05$) the emulator deviates by up to 5\%. The models with $f_{R_0} < 10^{-5}$ are predicted at sub-percent accuracy (corresponding zoom-in panel shown at the top). }
    \label{fig:emu}
\end{figure}

In this section we show the comparison of the emulator with COLA results. Our trained emulator is tested on parameter values within the limits of our SLH design, but not at the exact cosmologies used to construct the emulator.

\begin{table}
\caption{Cosmological parameters for the six additional COLA test simulations.}
\sisetup{round-mode=places}
\centering
{
\begin{tabular}{c S[round-precision=3]
S[round-precision=3] S[round-precision=3]
S[round-precision=3] 
S[round-precision=3] }

Model & $\Omega_m h^{2}$  & $n_{s}$ & $\sigma_{8}$ & log {\it (f$_{R_{0}})$} & $n$\\
\hline
T00  &  0.12466667 &  0.95666667 &  0.86     & -6.66666667 &  0.        \\
T01    &  0.13633333 &  1.02333333 &  0.83333333 & -4.    &  2.13333333\\
T02    &  0.15033333 &  0.97 &  0.82      & -6.13333333 & 3.2       \\
T03  &  0.13166667 &  0.89  &  0.79333333 & -5.86666667 &  2.93333333\\
T04    &  0.12933333 &  0.98333333 &  0.80666667 & -6.4  & 3.73333333\\
T05    & 0.127  & 1.05    &  0.74  &     -4.26666667 & 0.26666667 \\

\hline
\end{tabular}%
\npnoround
}
\label{T_extra}
\end{table}

In Fig.~\ref{fig:emu} we show the predictions of the emulator compared to an additional test set of six COLA simulations. The power spectrum ratios for the additional cosmologies  ($T00-T05$, parameters are given in Table~\ref{T_extra}) are randomly chosen within the allowed parameter ranges specified in Eqns.~(\ref{equ:params1}) -- (\ref{equ:params2}). The gray lines in the figure show all power spectrum ratios used to build the training set. The lower panel in the figure shows the relative error of the emulator output to the corresponding COLA results. The relative difference is within 5\% for our six test models. Sub-percent level accuracy is observed for models with $f_{R_0} < 10^{-5}$. The variation in accuracy corresponds to the sampling density of power spectrum ratios in the training set. For instance, the test simulation $T01$ with  $\log(f_{R_0}) = -4$ is at the very edge of our SLH design used for the training, and the $T05$ model is also close to the edge of the training design, with $\log(f_{R_0}) = -4.267$. By design, we focus more on exploring the modified gravity sector of $f_{R_0} < 10^{-5}$ that corresponds to smaller boosts in the matter power spectra. We stress that the required accuracy in the estimation of the power spectrum is 1-2 percent~\cite{Huterer:2004tr}, down to $k\sim 1 h$Mpc$^{-1}$, which we satisfy for most of the target parameter space.

\subsection{Comparison with N-body Simulations}\label{sec:nbodycomp}

\begin{figure*}

	\includegraphics[width=1.\textwidth]{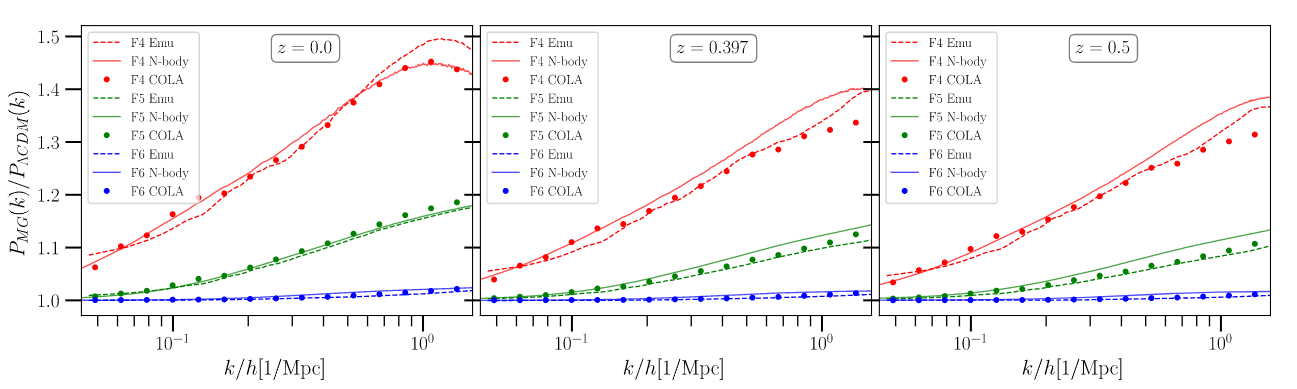}
    \caption{Comparison of emulator predictions to our  corresponding COLA simulations and the N-body simulations of Ref.~\citep{Cautun:2017tkc} for the F4 (red), F5 (green) and F6 (blue) models, respectively. Each panel represents a different redshift: $z=0$ (left), $z=0.397$ (middle) and $z=0.5$ (right). Solid lines are from the three N-body simulations, dashed lines are from corresponding emulator (with $z$-interpolation) output trained and the data points are COLA simulations performed for these models. The COLA and N-body runs correspond to a cosmology of $h=0.697$ while the emulator was built using $h=0.67$. The ratios of the power spectra are minimally impacted by these different choices.}
    \label{fig:nbody comp}
\end{figure*}

Next, we present a comparison of the MG emulator predictions against state-of-the-art N-body simulations for the Hu-Sawicki model. These are the Extended LEnsing PHysics using ANalaytic ray Tracing \verb|ELEPHANT| simulations \citep{Cautun:2017tkc}, that were performed with a modified version of the \verb|RAMSES| code,   
the \verb|ECOSMOG| module \citep{1475-7516-2012-01-051,Bose:2016wms}. Power spectra at various redshifts have been measured for the Hu-Sawicki case of $n=1$ and three variations of $|\bar{f}_{R_0}|=\{10^{-6},10^{-5}, 10^{-4}\}$. We refer to these as F6, F5 and F4, respectively, in the following discussion. The simulations evolved $1024^3$ dark matter particles, in a simulation volume of $V_{box}=(1024 h^{-1}$Mpc)$^3$ and a $\Lambda$CDM cosmology specified by the following parameters:
$\{\Omega_{m},\Omega_{\Lambda},h,n_s,\sigma_8, \Omega_b\}=\{0.281,0.719,0.697,0.971,0.82, 0.046\}$. For each model, five different random realizations are available. More details about these simulations can be found in Ref.~\citep{Cautun:2017tkc}.

In Fig.~\ref{fig:nbody comp}, we compare the predictions from our emulator to the full N-body simulations from Ref.~\citep{Cautun:2017tkc} for the F4, F5 and F6 cases at three different redshifts: $z=0$, $z=0.397$ and $z=0.5$, which span the redshift range in which predicted MG signals are more pronounced. 
Furthermore, in order to independently illustrate the accuracy of our training set, we show the ratio $P_{MG}(k)/P_{\Lambda CDM}(k)$ obtained from COLA simulations that we additionally performed, separately from our design, for these scenarios. COLA is found to recover the simulated ratios at (better than) $1\%$ level of accuracy in the (F6) F5 case for all redshifts, in agreement with previous findings in the literature \citep{Valogiannis:2016ane}. For the F4 model, the agreement is still similarly good at $z=0$, but worsens progressively for higher redshifts, with COLA underestimating the predicted deviation. This behavior is most likely attributed to the approximate MG screening implementation \citep{Winther:2014cia} used in COLA, which is known to underestimate the ratio particularly in cases that deviate substantially from GR (such as F4) and at high $z$. 
We stress that this tendency is not detrimental, since the deviations typically predicted by cases such as F4 are of substantial magnitude \footnote{The screening mechanism effectively fails for such large values of $f_{R0}$}, and effectively ruled out by observations \cite{Lombriser2014,Burrage2018}. We do however choose to include larger values of $f_{R0}$ in our target range, to
enable the exploration of the linearized regime of the Hu-Sawicki MG models. 

Our emulator is found to successfully recover the target ratios for F5 and F6 at a very similar level of accuracy as the COLA method, for all three redshifts. The predictions are accurate even for the F4 corner case.  
All of the above findings are consistent with the levels of accuracy previously found by the accuracy tests shown in Sec. \ref{sec:holdout}. 

We end this section by clarifying that the COLA and full N-body simulations shown in Fig.~\ref{fig:nbody comp} correspond to $h=0.697$, whereas the emulator predictions were generated from a COLA training set that assumed $h=0.67$, in agreement with the Planck constraints \citep{Ade:2013zuv, Ade:2015xua}. We have carefully checked and confirmed that this small inconsistency leads to negligible errors in the emulated ratio for all cases and redshifts of our design. As a result, the direct comparison in Fig.~\ref{fig:nbody comp} is indeed meaningful, despite the different underlying values of $h$ assumed.

%% file: application_sec.tex
We present three preliminary applications for the emulator developed in this paper. Using the power spectrum ratio as the summary statistic we first perform a parameter sensitivity analysis. Second, we use the emulator for Fisher forecasting. Finally, results from an MCMC run for a fiducial cosmology are shown. The evaluation time for an emulator prediction is less than 0.001 seconds per computation on an Intel Core i5 Processor, delivering a massive speed-up over numerical calculation using COLA, which typically takes about an hour on similar computational hardware. This is particularly important for our third application, where our GP emulator is implemented in the MCMC likelihood calculation which requires a very large number of accurate predictions for the power spectrum ratio.

\subsection{Parameter Sensitivity Analysis}

In this section we investigate the effect of different cosmological parameters on the power spectrum ratio $P_{MG}(k)/P_{\Lambda CDM}(k)$ using the emulator. For this study, we choose a base model at redshift $z=0.0$. The base model is evaluated at parameters shown in Equation \ref{eq:fid_cosmo_params}:

\begin{equation}\label{eq:fid_cosmo_params}
\begin{split}
    \Omega_m h^2 &= 0.142, \\
    n_s &= 0.967, \\
    \sigma_8 &= 0.816, \\
    \log(f_{R_0}) &= -5, \\
    n &= 1.0
\end{split}
\end{equation}

\begin{figure*}
	\includegraphics[width=0.75\textwidth]{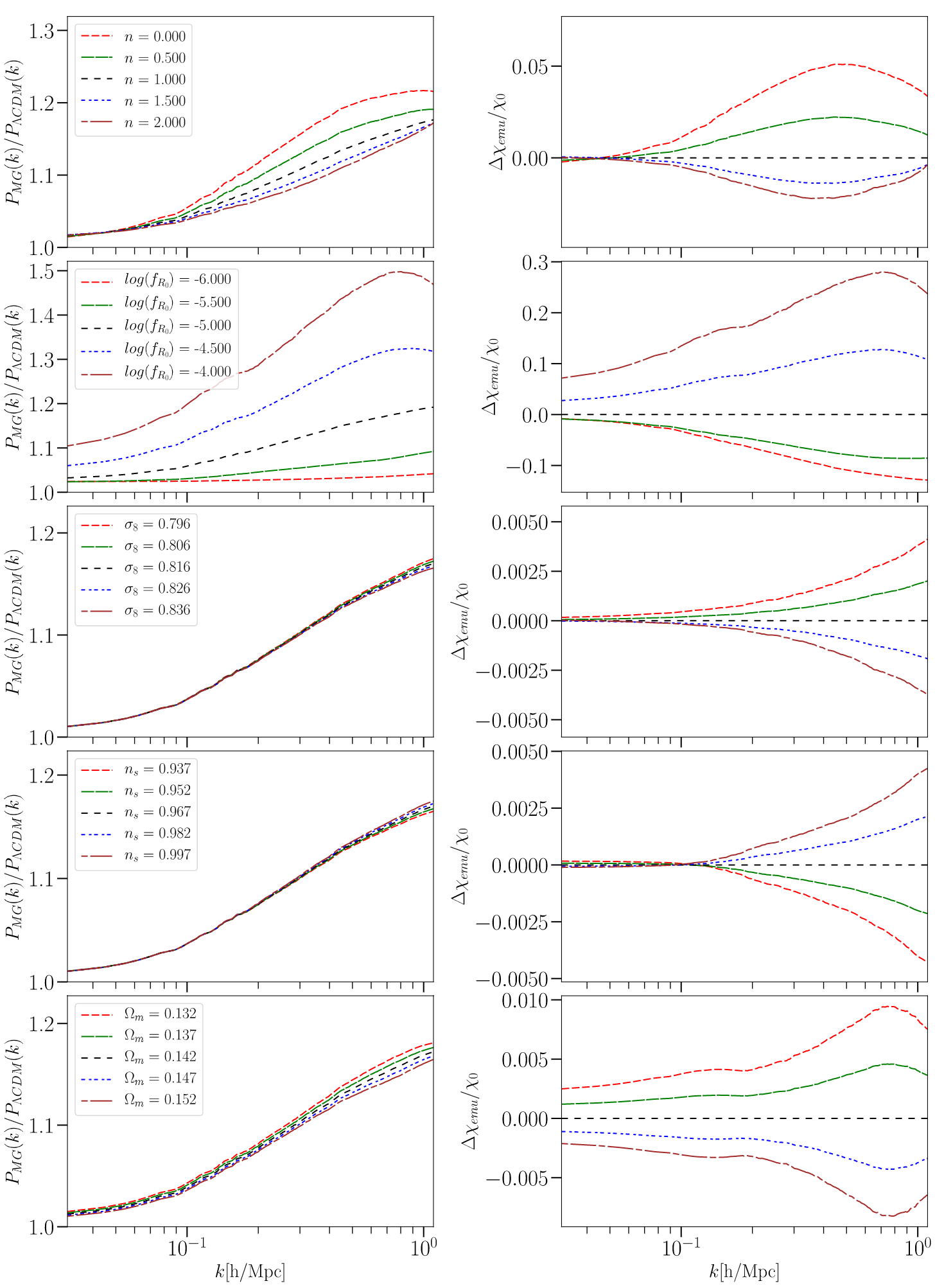}
    \caption{Results for the sensitivity analysis for the power spectrum ratios using emulator predictions. The impact of five cosmological parameters is investigated. 
    The left panels show the emulator outputs from linearly varying one parameter while keeping the rest at the base cosmology. The right panels show the relative variation of the the power spectrum  ratios $\Delta \chi_{emu} / \chi_0$ with respect to the emulator output $\chi_0$ at base cosmology parameters shown in Equation \ref{eq:fid_cosmo_params}. From top to bottom the parameter sensitivity analysis with respect to $n$, $\log(f_{R_0})$, $\sigma_8$, $n_s$ and $\Omega_m h^2$ is shown respectively. Note that the scales on the $y$-axes for the relative variation differ by up to two orders of magnitude. Given the wide range we allowed for $\log(f_{R_0})$ it is not surprising that its impact is by far the largest on the power spectrum ratio.}
    \label{fig:sensitivity}
\end{figure*}

We then measure the sensitivity of the power spectrum ratio to changes in one cosmological parameter at a time while keeping the others at their base values. We stress that the MG parameters span a much broader range than the $\Lambda$CDM parameters and therefore we expect a much larger impact on the power spectrum ratio for varying $\log(f_{R_0})$ and $n$.
The results are presented in Fig.~\ref{fig:sensitivity}.

Primarily, and in agreement with the literature \citep{Hu:2007nk,Zhao:2010qy}, we observe that $\log(f_{R_0})$ has the highest 
impact on the matter power spectrum ratios, showing up to 40 percent variation just around $ 0.3 h$Mpc$^{-1} \lessapprox  k \lessapprox  1 h$Mpc$^{-1}$ for the range $-6 \leq \log(f_{R_0}) \leq -4$. Larger $\log(f_{R_0})$ results in enhancement of the power spectrum ratios, due to the progressive weakening of the screening mechanism, and the increase is monotonous across the full range of testing cosmologies. The analysis also suggests that with $\log(f_{R_0}) > -5$ the modified gravity matter power spectrum $P_{MG}(k)$ is up to 50 percent higher than $P_{\Lambda CDM}(k)$. This shows that models in this range are disfavored by current data, which is consistent with previous studies constraining $f(R)$ models using available data sets, see for example Refs.~\cite{Lombriser2014,Burrage2018}.

The second highest contribution to the changes in the summary statistic is due to the second Hu-Sawicki MG parameter, $n$. The variation is under 10\% from the base cosmology for the range $0 \leq n \leq 2$. However, the peak of the departure occurs at slightly larger length scales, between $ 0.1 h$Mpc$^{-1} \lessapprox  k \lessapprox  0.9 h$Mpc$^{-1}$. For $ k \lessapprox 1 h$Mpc$^{-1}$, larger values of $n$ cause higher suppression of the matter power spectrum ratios. This trend reverses beyond $ k \gtrapprox 1 h$Mpc$^{-1}$. Unfortunately, there are currently no N-body Hu-Sawicki simulations for $n \neq1$ available in the literature, and thus a thorough comparison on nonlinear scales is not possible. Nevertheless, our observed large-scale trend seems to be in qualitatively agreement with the linear considerations in Ref.~\citep{Hu:2007nk}. Even though we do not expect the behavior of the model to be substantially different for $n \neq1$, we defer a detailed study of the emulator accuracy for such values to future work, when corresponding N-body simulations become available.

Finally, the relative change of the emulator output when varying the $\Lambda$CDM parameters is more restricted. Both $\sigma_8$ and $n_s$ reveal sub-percent variation within their respective ranges of  $0.796 \leq \sigma_8 \leq 0.836$ and  $0.937 \leq n_s \leq 0.997$. Moreover, both their peak departures from the base model occur at $ k \gtrapprox 1 h$Mpc$^{-1}$, where the accuracy of the training COLA simulations with respect to full N-body simulations reduces at higher redshifts. Also, the variations of $\sigma_8$ and $n_s$ beyond  $ k \lessapprox 0.1 h$Mpc$^{-1} $, are opposite of each other, i.e, increasing $\sigma_8$ reduces the $P_{MG}(k)/P_{\Lambda CDM}(k)$ ratio, whereas $n_s$ has the opposite effect. Variations of the final emulator parameter, $\Omega_m h^2$, lead to a monotonic change in the matter power spectrum ratios, where increasing the values from 0.132 to 0.152 shows a decrease in $P_{MG}(k)/P_{\Lambda CDM}(k)$. This reduced sensitivity of the ratio with respect to variations of the $\Lambda$CDM parameters (relative to the response of MG parameters) was also found using the fitting formula approach in Ref.~\cite{Winther2019}.

\subsection{Fisher Forecasting}\label{Fishersubsection}

The likelihood $\mathcal{L}(\chi|\theta)$ is defined as the probability distribution function of an observed summary statistic $\chi$ for a given model with parameters $\theta$. The emulator output $\chi_{\rm{emu}}(k; \theta) = P_{MG, \rm{emu}}(k)/P_{\Lambda CDM, \rm{emu}}(k)$ at a given redshift itself can be considered as the forward model in the computation of the likelihood. In the case of the observed power spectrum ratio $\chi_{ \rm{obs}}(k) = P_{MG, \rm{obs}}(k)/P_{\Lambda CDM, \rm{obs}}(k)$ for a set of cosmological parameters $\theta=\{\Omega_m h^2, \sigma_8, n_s , \log(f_{R_0}), n  \}$, the likelihood is computed using Equation \eqref{eq:lnlike} assuming it is of a Gaussian form: 

\begin{multline}
\label{eq:lnlike}
\log\mathcal{L}(\chi|\theta) \propto -{\frac {1}{2}}\left[\chi_{\rm{emu}}(k; \theta)-\chi_{ \rm{obs}}(k)\right] \widehat{{\mathit { C_{ij} }}^{-1}} \\
\left[\chi_{\rm{emu}}(k; \theta)-\chi_{ \rm{obs}}(k)\right]^{\mathrm {T} }.
\end{multline}

\begin{figure}
	\includegraphics[width=0.48\textwidth]{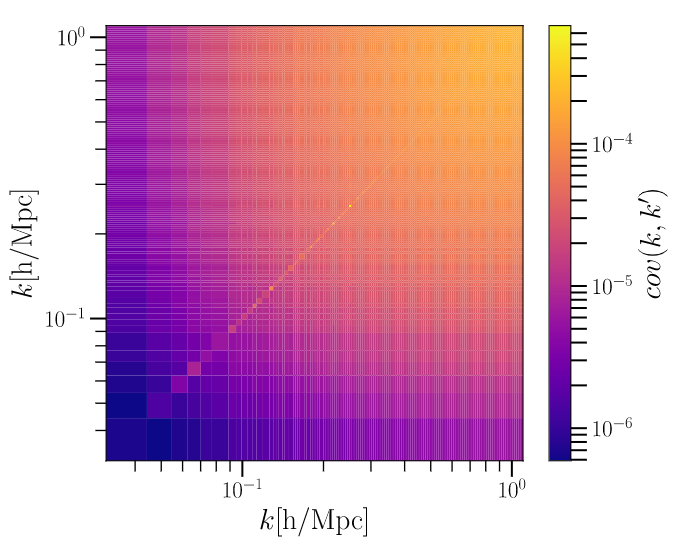}
    \caption{Covariance matrix for the ratio $P_{MG}(k)/P_{\Lambda CDM}(k)$ obtained from 2000 COLA realizations for our fiducial cosmology parameters $\theta_{fid}$ at redshift $z= 0.02$ as an example. The likelihood obtained using this covariance matrix is used for both Fisher analysis and posterior estimation of the cosmological parameters via MCMC sampling.}
    \label{fig:cov}
\end{figure}

We construct a mock data set for the power spectrum ratios with an associated mean, $\chi_{ \rm{obs}}(k)$, and a covariance matrix $C_{ij}$, which captures the effects of cosmic variance. This data set is computed using 2000 COLA simulation realizations run at a fiducial cosmology
$\allowbreak \theta_{fid} \equiv \{\Omega_m h^2 = 0.142, n_s = 0.967, \allowbreak \sigma_8 = 0.816, \allowbreak \log(f_{R_0}) = -5, \allowbreak n = 1.0\}$,
shown in Figure \ref{fig:cov}. The unbiased estimator for the inverse covariance matrix $\widehat{{\mathit { C_{ij} }}^{-1}}$ is then computed using Equation \eqref{eq:cov_in} given by Ref.~\cite{2007A&A...464..399H}. This correction accounts for the number of COLA simulations used ($N=2000$) and the size of the data vector $D$, which depends on our range of wavenumbers used in calculating the likelihood: 

\begin{equation}
\label{eq:cov_in}
\widehat{{\mathit { C_{ij} }}^{-1}} = \frac{N-D-2}{N-1} { C_{ij}}^{-1}.
\end{equation}

The box size, mass resolution and other simulation specifications for these additional COLA simulations are the same as the ones listed in Table ~\ref{T_sims}. We choose a single redshift of $z= 0.02$ for this analysis. 

The Fisher information matrix assesses how well a cosmological parameter can be inferred from a summary statistic. It is defined in terms of the likelihood $\mathcal{L}$ of the data in the following equation: 

\begin{equation}
    F_{ij} = \left\langle \frac{\partial^2 log \mathcal{L}}{\partial \theta_i \partial \theta_j} \right\rangle
    \label{eq:fisher}
\end{equation}

\begin{figure*}
	\includegraphics[width=0.90\textwidth]{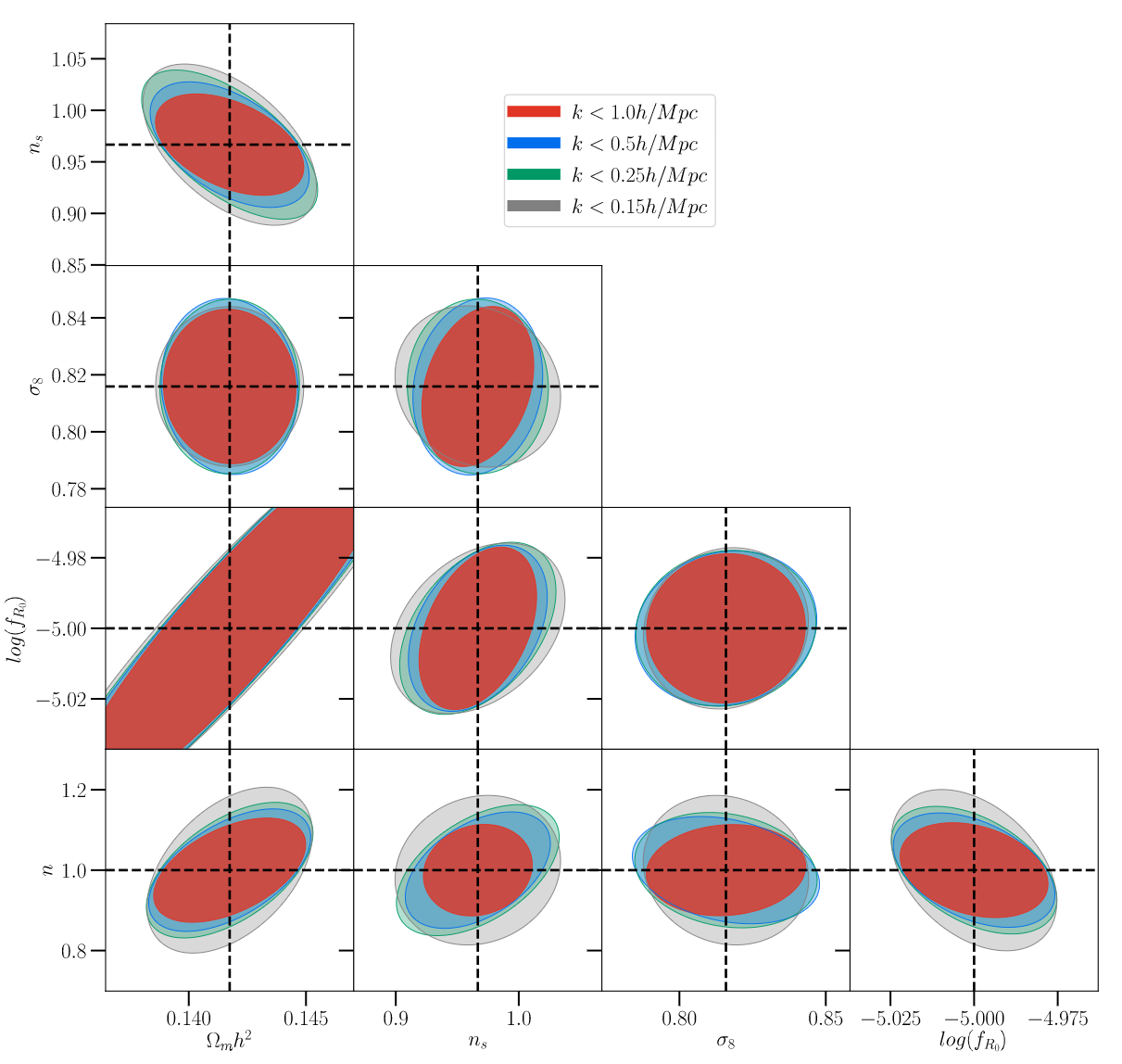}
    \caption{Confidence ellipses from the Fisher information matrix $F_{ij}$ for the cosmological parameters considered and using the emulator to provide the power spectrum ratio information.  Each panel shows the covariance between the $\Lambda$CDM and MG parameters. The thresholds for the wave-numbers used in the computation of the likelihood, i.e., $k_{max} = 0.15h$Mpc$^{-1}$, 0.25$h$Mpc$^{-1}$, 0.5$h$Mpc$^{-1}$ and $ 1.0 h$Mpc$^{-1}$ correspond to the four 1-$\sigma$ level confidence ellipses in each panel. The dashed lines in each panel correspond to the fiducial cosmological parameters $\theta_{fid}$. }
    \label{fig:ellipses}
\end{figure*}

The Fisher matrix can be calculated directly using the emulator, either using numerical derivatives, or using the analytical derivatives of Gaussian Processes propagated through the emulator. We choose the former method and calculate the second order partial derivatives numerically. The corresponding confidence ellipses are shown in Figure~\ref{fig:ellipses}. The numerical derivatives are evaluated with multiple step sizes to ensure consistency.

The Fisher information obtained from the power spectrum ratio analysis reveals correlations between MG parameters and $\Lambda$CDM parameters. 
We present results using 4 different wavenumber thresholds ($k_{max} = 0.15h$Mpc$^{-1}$, 0.25$h$Mpc$^{-1}$, 0.5$h$Mpc$^{-1}$ and $ 1.0 h$Mpc$^{-1}$) corresponding to increasing scales of nonlinear growth of structure. 
The tightest constraints for forecasts across all parameters are achieved when including nonlinear scales up to $k_{max} = 1.0 h$Mpc$^{-1}$, and the constraints weaken with decreasing nonlinearity. This is expected, given that including more modes gives us increasingly more information about the underlying cosmological parameters (including the MG parameters), resulting in increased constraining power. 

It should be noted that the aim of this exploratory set-up is not to perform a thorough study of forecasted MG constraints, but rather to highlight the diverse applications of our emulator. As a result, these constraints only provide an approximate view. For a more realistic approach, one would need to combine the emulated ratio with a prediction for $P_{\Lambda CDM}(k)$ (e.g. from the \verb|CosmicEmu|), as well as fold in prescriptions for the galaxy bias and/or redshift-space distortions \citep{Valogiannis:2019nfz}, in order to obtain a prediction for the observed galaxy power spectrum in MG (as for example was done in \citep{Winther2019}). This step goes beyond the scope of this paper, but will be studied in detail in future work.
As a consequence, our constraints are not directly comparable to the ones obtained by the fitting formula approach in \citep{Winther2019}.

\subsection{Parameter Constraints Using MCMC}\label{sec:mcmc}

The speed and precision of our emulator enables quick parameter inference using traditional Bayesian inference schemes like Markov Chain Monte Carlo (MCMC) methods. We illustrate the application of such emulator-based posterior estimation for mock data, the same as shown in Sec.~\ref{Fishersubsection}. The Gaussian likelihood is computed via Equation \eqref{eq:lnlike}, using the emulator and the mock covariance matrix shown in Fig. \ref{fig:cov} from 2000 COLA realizations for the fiducial cosmology parameters $\theta_{fid}$. Our aim is to demonstrate the recovery of the original parameter values within appropriate error margins. In addition, the exploration of the posterior distribution allows us to study the covariance between the parameters that arises from the data vector and sensitivity of the parameters. 

\begin{figure}
	\includegraphics[width=0.48\textwidth]{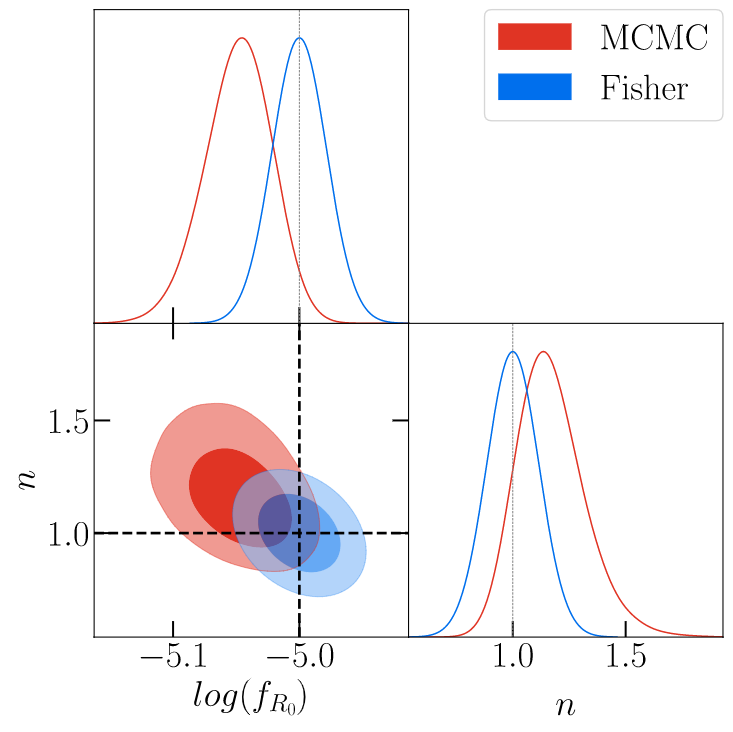}
    \caption{Bayesian posterior distribution for $\{\log(f_{R_0}, n)\}$ obtained using MCMC sampling. The data vector, emulator output and likelihood computation are carried out for wave-numbers $k < 1.0 h$Mpc$^{-1}$. The two shades of red correspond to 1-, 2-$\sigma$ confidence intervals of the posterior. Dotted lines are the target cosmologies of the mock data, $\log(f_{R_0} = -5$ and $ n = 1$. The constraints are obtained from uniform priors in the ranges $-8 \leq \log(f_{R_0}) \leq -4$ and  $0 \leq n \leq 4$. The blue contours are the 1- and 2-$\sigma$ confidence ellipses from the Fisher information constraints for the same $k$-range at the fiducial cosmology,  showing the consistency between the two applications in space and direction of correlation.}
    \label{fig:mcmc2}
\end{figure}

We execute two MCMC sampling schemes. First with fixed values for the $\Lambda$CDM parameters at $\{\Omega_m h^2 = 0.142, n_s = 0.967, \allowbreak \sigma_8 = 0.816 \}$, we evaluate the posterior distributions of the 2 MG parameters. The logarithmic value for $f_{R_0}$ is chosen to reduce the dynamical range of the MCMC sampling, and a flat prior in the range $-8 \leq \log(f_{R_0}) \leq -4$ is selected. For $n$ we also use a flat prior within the limits of our experimental design: $0 \leq n \leq 4$. While the choice of priors could be more stringent, we select such uninformative priors spanning the entire parameter range of the emulator in order to avoid obtaining  prior-dominated constraints for the MG parameters. The likelihood $\mathcal{L}(\chi|\theta)$ of the mock data vector is computed up to $k < 1.0 h$Mpc$^{-1}$. 

We utilize an MCMC approach to approximate the posterior distribution using an affine-invariant ensemble sampling \cite{Goodman2010} method implemented in the \verb|emcee| \cite{emcee} sampler. An ensemble of 300 walkers or Markov chains with 3000 evaluations or steps per walker only takes about 4000 seconds on a single processor, hence enabling quick explorations of the posterior space of the parameters of interest. The final constraints obtained were $\theta_{MCMC} \equiv \{\log(f_{R_0}) = -5_{-0.04}^{+0.05}, n = 1.18_{-0.20}^{+0.20}\}$, as shown in the top panel of Fig.~\ref{fig:mcmc2}, and a recovery of parameters within the 2-$\sigma$ margin of error is found. While the 2000 realizations for fiducial covariance matrix have enabled tight parameter constraints via MCMC sampling compared to the broad priors, overcoming the offset within 2-$\sigma$ margin of error from the fiducial targets may require further reduction in noise. The emulator outputs at the $\theta_{fid}$ and $\theta_{MCMC}$ are matched well with the $P_{MG}(k)/P_{\Lambda CDM}(k)$ ratio of the fiducial cosmology simulations. Moreover, the direction of the covariance from the Fisher forecasts in Section~\ref{Fishersubsection} also matches with the contours of the posteriors. The consistent results are shown in the bottom left panel of Fig.~\ref{fig:mcmc2}. 

\begin{figure*}
	\includegraphics[width=0.90\textwidth]{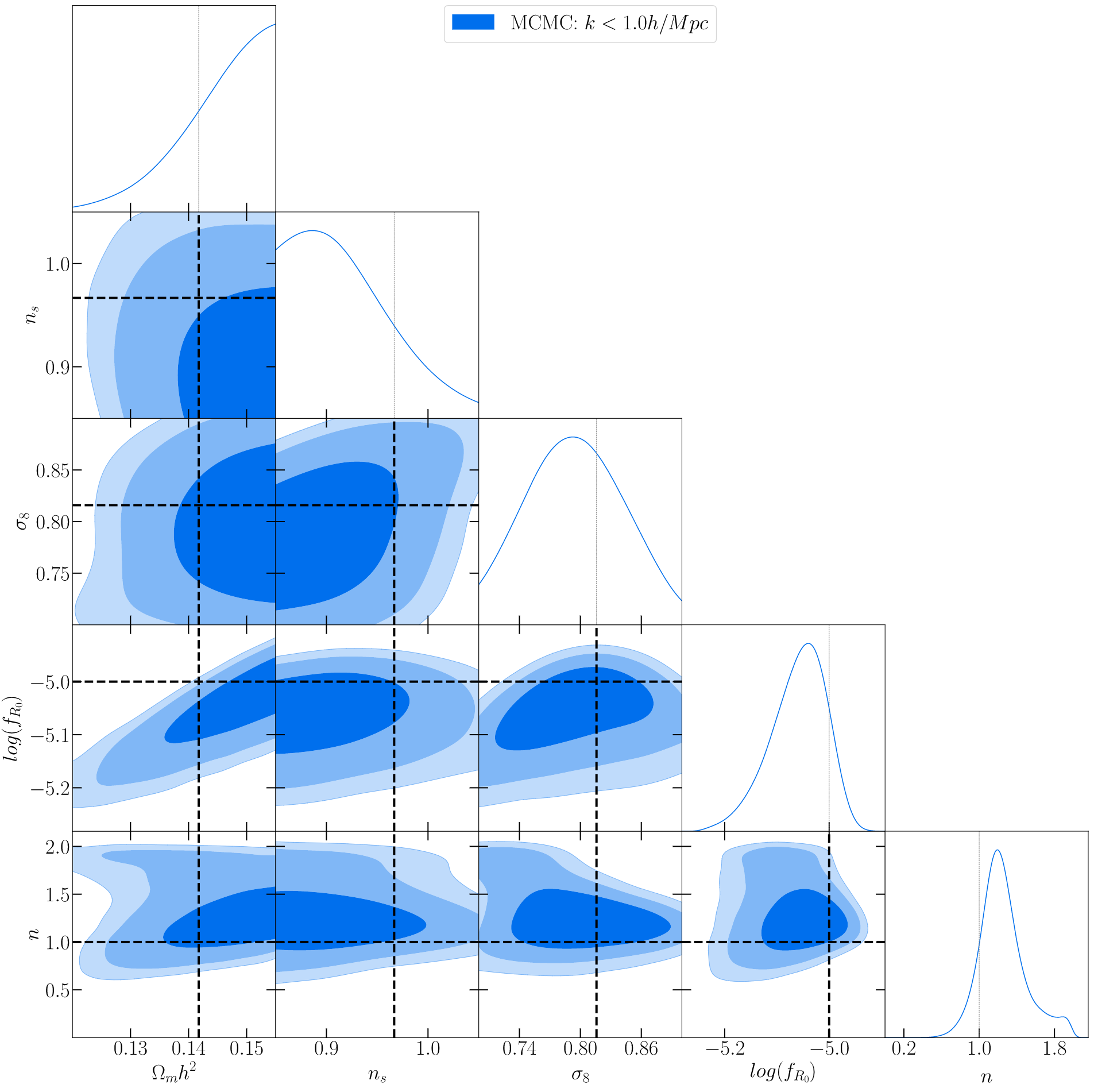}
    \caption{Posterior distribution evaluated from an MCMC sampling for five cosmological parameters. The data vector, emulator output and likelihood function are computed for wave-numbers $k < 1.0 h$Mpc$^{-1}$. The three shades of blue correspond to 1-, 2- and 3-$\sigma$ confidence intervals. Dotted lines are the target cosmological parameters of the mock data set $\theta_{fid}$. It is very satisfactory to observe that 
    the target cosmological parameters fit well within the 2-$\sigma$ confidence intervals, despite the relatively limited constraining power of the ratio of power spectra when used with no additional cosmological information.}
    \label{fig:mcmc5}
\end{figure*}

In addition, we also sample the combined posterior distribution of the $\Lambda$CDM  parameters \{$\Omega_m h^2, \sigma_8, n_s$\} and the $f(R)$ Hu-Sawicki model parameters \{$\log ( f_{R_0}), n$\}  using an MCMC approach. The fiducial cosmology $\theta_{fid}$, the covariance information and the likelihood function remain the same, with a $k$-range limited to $k < 1.0 h$Mpc$^{-1}$ for the computation of the likelihood. The priors are again chosen to be flat within the full range covered by the emulator. The corresponding posterior distribution functions are shown in Fig.~\ref{fig:mcmc5}, along with the target cosmological parameters corresponding to the mock data set. The final constraints corresponding to the MCMC sampling are $\allowbreak \theta_{MCMC} \equiv \allowbreak \{\Omega_m h^2 = 0.15_{-0.01}^{+0.01} , \allowbreak n_s = 0.91_{-0.04}^{+0.06}, \allowbreak \sigma_8 = 0.79_{-0.05}^{+0.05}, \allowbreak \log(f_{R_0}) = -5.05_{-0.05}^{+0.05}, \allowbreak n = 1.22_{-0.19}^{+0.25}\}$. 

The MCMC sampling of the parameter space is restricted to the parameter range in our SLH design. Especially for $\Omega_m h^2$, $\sigma_8$ and $n_s$, this results in a partial coverage of the posterior distribution. This is displayed in the individual panels of Fig.~\ref{fig:mcmc5}, where the posteriors are  limited to the parameter range of the emulator. These constraints would be tightened using a joint analysis (with CMB data, for instance). In this study we restrict our attention to explorations of the constraining power of $P_{MG}(k)/P_{\Lambda CDM}(k)$ alone, and reserve combined analyses for future studies. We additionally note that MCMC constraints for the HS model were also presented in \citep{Schneider_2020, Bose:2020wch}, but using the power spectrum. For this reason, our results are not directly comparable with the ones in these studies.

The 1, 2 and 3-$\sigma$ contours for the MG parameters \{$\log ( f_{R_0}), n$\} are within the range of our experimental design, as seen in both Fig.~\ref{fig:mcmc2} and Fig.~\ref{fig:mcmc5}. We note that the constraints obtained from sampling a parameter space of just 2 parameters are tighter than those for 5 parameters. Posterior estimation restricted to fewer parameters removes possible degeneracies, resulting in reduced sampling space.

The original parameters corresponding to the mock data vectors are recovered in both the sample MCMC runs, indicating that the boost in the matter power spectra is a powerful statistic for constraining MG parameters, especially when coupled with robust emulators as presented in this work. In both posterior approximations, the chosen prior probability distributions are highly uninformative, i.e., they are uniform with coverage of the entire training limits. Posterior contours in Figure \ref{fig:mcmc2} display a shift from the fiducial values, and a partial coverage of posterior distribution is seen in Figure \ref{fig:mcmc5}.  
These constraints on the cosmological parameters would be improved either using a tomographic or multi-probe analysis, and we reserve these for a future study of modeling cosmological observables and associated forecasting in MG scenario.

%% file: conclusion_sec.tex
In this paper, we present an emulator for efficiently predicting the enhancements in the nonlinear matter power spectrum due to beyond-GR effects. This emulator is based on the $f(R)$ Hu-Sawicki model, a MG candidate prioritized for further studies in near-future Stage-IV surveys such as the LSST and Euclid. We aim at simplifying the tasks of model selection and cosmological parameter inference by the use of fast and robust generation of the power spectrum ratio $P_{MG}(k)/P_{\Lambda CDM}(k)$. The emulator
provides estimations across the redshift range of $ 0 < z < 49$, for a combination of cosmological parameters \{$\Omega_m h^2, \sigma_8,n_s, \log ( f_{R_0}), n$\}. This way, we enable the full exploration of the parameter space that defines the Hu-Sawicki model.

Our emulator is trained on simulations produced by the efficient COLA method, which effectively captures the chameleon screening mechanism through a phenomenological thin-shell factor attached to the scalar field Klein-Gordon equation. The matter power spectrum ratios extracted are computed by running two consistent COLA simulations at each training point -- one for MG and the other for the corresponding $\Lambda$CDM scenario, reducing the effect of cosmic variance while at the same time highlighting the effects of MG. Our fully trained emulator is validated on additional cosmologies within our target parameter space and is found to achieve sub-percent levels of accuracy for models with $f_{R_0} < 10^{-5}$ and up to $5 \%$ agreement when $f_{R_0} > 10^{-5}$. The computation time is less than 0.001 seconds, delivering thus a massive speed-up by 6 orders of magnitude compared to the COLA simulations.

In order to explore and validate the diverse capabilities of our emulator, we further proceed to utilize its predictions for three preliminary applications. 
First, we perform a sensitivity study of the target summary statistic with respect to the variation of the five cosmological parameters. We find that the power spectrum ratio exhibits the highest sensitivity for the MG parameter 
$\log(f_{R_0})$ followed by the $n$ parameter, in agreement with previous studies \citep{Hu:2007nk,Zhao:2010qy}. Next, we produce constraints around a fiducial cosmology using Fisher forecasting as well as MCMC parameter inference. The confidence contours obtained are consistent between the two methods and also consistent with the corresponding analytical expressions. The emulated ratio is thus found to enable accurate constraints for both MG parameters as well as the values of the background $\Lambda$CDM cosmological parameters. 

 We conservatively limit the use of the emulator to $k \leq 1.0 h$Mpc$^{-1}$ throughout the analysis of this paper. Our tests shows that the COLA prescription agrees with available N-body simulations within a relative error of 5 $\%$ up to $k \sim 1.0 h$Mpc$^{-1}$, and hence we advocate the use of this emulator only up to this limit. 

We also advise the reader to exercise caution in extrapolating the emulator beyond the limits of cosmological parameters and redshifts used in the experimental design. Gaussian Processes, like any interpolation schemes, may give estimates with large extrapolation uncertainties beyond the training range. We note again that our tests on additional COLA simulations show that models with $\log(f_{R_0}) < -5$ agree within 1\%, while the estimation error rises up to 5\% for larger values of $\log(f_{R_0})$. We also note that the emulator is trained on one realization per cosmology, with just 50 training points. With a large training sample with better sampling and larger number of realizations per cosmology, the relative error with respect to COLA simulations is naturally expected to reduce.

In addition to enabling an efficient exploration of the deviations predicted by MG, our emulator also serves as a stepping stone to allow a broad portfolio of future applications. Through a simple multiplication by the $\Lambda CDM$ power spectrum (from emulators like \verb|CosmicEmu|), the emulated power spectrum ratio can straightforwardly provide a prediction for the MG nonlinear matter power spectrum itself.
The latter can then also be utilized to incorporate the effects of galaxy bias and Redshift Space Distortions (RSD) in MG (for example as in Ref.~\citep{Valogiannis:2019nfz}), which are crucial for comparing with observations, or to enable obtaining MG constraints from weak lensing cosmic shear measurements. Such predictions should however take into account the accuracy and sensitivity of the estimators at corresponding length scales, and we reserve this for future studies. Furthermore, the emulator will be a useful standardized routine to be included in the Core Cosmology Library (CCL \footnote{\url{https://github.com/LSSTDESC/CCL}}, \cite{2019ApJS..242....2C}) in order to calculate basic cosmological observables, specifically for Rubin Observatory LSST analyses. We also intend in a follow-up work to employ the emulator in providing forecasts using a detailed joint probes cosmological parameters analysis. Last but not least, we plan to expand our emulator's capabilities in order to incorporate the effects of massive neutrinos, and also to support the more general class of Horndeski MG models prioritized by our collaboration. The necessary modifications to achieve these steps with the efficient COLA approach are already underway. The combined outcome of these efforts will be a diverse set of tools that will allow efficient and reliable explorations of the broad spectrum of beyond-$w(z)$CDM candidates.

In this decade, precise cosmological observations will offer a  unique opportunity to constrain beyond-$w(z)$CDM models at an unprecedented level of accuracy. Emulators like the one developed in this work will be essential and necessary to perform cosmological analyses with accurate and fast theoretical predictions in the nonlinear regime, and to take full advantage of the formidable data expected from Rubin Observatory LSST, DESI, Euclid, SPHEREx and Roman Space Telescope.

%% file: smooth_appendix.tex
\label{sec:smooth_appendix}

The Savitzky-Golay smoothing filter  \cite{Savitzky1964, Press1990} performs convolution operations on adjacent data points with a polynomial function, which gives us the effect of smoothing the input dataset. The window size and the order of the polynomial are the two parameters that specify the smoothing operation. Equation \eqref{eq:savgol} shows the value of $j$-th bin of the smoothed power spectrum ratio $\chi(k)$:

\begin{equation}
\label{eq:savgol}
\chi(k_j) =\sum _{i={\tfrac {1-m}{2}}}^{\tfrac {m-1}{2}}C_{i}\, \frac{P_{MG}(k_{j+i})}{P_{\Lambda CDM}(k_{j+i})}.
\end{equation}

The individual convolution coefficients $C_i$ are defined by the analytical expression given in Ref.~\cite{Savitzky1964}. Within every window, a polynomial (of order $p$) is fitted, which provides a smoothing effect of the input dataset. The window size $m$ is the number of data points chosen for individual regression.

The tuning of the two free parameters $m$ and $p$ depends on the largely on the type of data and the desired level of smoothing. These parameters are hand-tuned for smoothing the power spectrum ratios, and checked for consistency with various choices of smoothing filters and window sizes. As the ratio of window width to polynomial order, $m/p$ increases the amount of smoothing increases. First, the window width was appropriately tuned to be effective against the noise and is set to be $m = 51$ points. For this window width, a third order polynomial ($p = 3$) is fitted. A polynomial of order $p>3$ would closely follow the undesired noise resulting from the single realizations of the COLA simulations. For a given polynomial order, decreasing the window length has a similar effect, i.e., while the bias decreases (the smoothing function closely follows the raw data power spectrum ratios from the COLA simulations), the estimation variance increases, resulting in an over-fitted smoothing function.

Smoothing near the boundaries requires additional considerations. The data points at the edges cannot be placed at the center of a symmetric window, hence the Equation \eqref{eq:savgol} is applied for ${\frac {m-1}{2}}\leq j\leq n_{bins}-{\frac {m-1}{2}}$ only. We use a separate treatment for smoothing at the boundaries, where the polynomial fitted to the windows near the edges of the data is used to evaluate the first and the last $m/2$ smoothed outputs of $\chi(k_j)$. The effect due to the edge effects is less significant at low-$k$ boundary (i.e., $k<0.17hMpc^{-1}$) since the raw $P_{MG}(k)/P_{\Lambda CDM}(k)$ is less noisy. On the higher end, this near-boundary effects smoothing at $k>2.35hMpc^{-1}$. Since our emulator outputs are restricted to $k\leq1hMpc^{-1}$, this does not effect our estimations directly. However, the consistency of the smoothing results across all the 213 bins (up-to $k = 3.5hMpc^{-1}$) is considered whilst tuning the free-parameters.

%% file: gp_appendix.tex
\label{sec:gp_appendix}

A parametric regression task \cite{fisher1992statistical} involves an estimation of finite number model parameters that fit the data. For $n$ training targets $\left\{y_1, \dots, y_n\right\}$ at training locations $\left\{x_1, \dots, x_n\right\}$, one may define a polynomial regression relationship and estimate the finite number of polynomial coefficients. With a frequentist approach, point estimates of these parameters can be estimated. Alternatively, a Bayesian approach treats these model parameters as probability distributions which are to be inferred using the training points. With these either a point estimation or distribution of a test target $y_{*}$ can be made at a new location $x_{*}$.

In contrast, Gaussian process \cite{Rasmussen2006} regression is a non-parametric approach i.e., one finds a distribution of possible functions $f(x)$ that are consistent with observed data. GPs are remarkably good Bayesian tools that perform regression tasks with associated uncertainties. Although the inference is fast, computational cost of GP regression can be very expensive and grows cubicly with the training set size and dimension. Due to this constraint, dimensionality reduction is usually performed on the training data, as is the case with our approach of building the emulator. We employ a data reduction technique called Principal Component Analysis (PCA) to reduce computational expenses during GP training. With a PCA decomposition (in Equation \ref{eq:pca}), the bases $\phi_i(k)$ are independent of the cosmological parameters, and only the weights $w_i(\theta)$ are used in GP interpolation. Thus the training locations are the set of parameters $\theta$ (tabulated in Appendix \ref{sec:params_appendix}) where 50 COLA simulations are computed, and the training targets are the corresponding weights $w_i(\theta)$.

With GP, we first assume that the joint probability distribution $p\left(f(x_1), \dots, f(x_n)\right)$ are jointly Gaussian with mean $\mu(x)$ and covariance $\mathbf{K}$, where the elements $\mathbf{K}_{ij} = k(x_i, x_j)$ with $k(x_i, x_j)$ being the covariance kernel. For simplicity, the GP prior can be defined using a zero mean and covariance as $p\left(f(x)\right) = GP(0, \mathbf{K})$. In our emulator construction, the weights are essentially sampled from this distribution, i.e.,  $w_i(\theta) \sim GP(0, \mathbf{K})$, where the covariance $\mathbf{K} = k(\theta, \theta{'})$.

The kernel function $k$ is usually selected depending on how smooth the function is expected to be. One popular choice is the squared exponential or the Radial basis function kernel: $k_{RBF}(x, x') = \sigma^2 \exp \left( -\frac{(x - x')^2}{2 \theta^2} \right)$ with hyperparameters $\left(\sigma^2, \theta\right)$ corresponding to the process variance and lengthscale, respectively. These hyper-parameters can be inferred based on the maximum likelihood or other more Bayesian techniques. Once optimized, the GP model has learned a distribution of functions that fits the training data. 

Using the GP assumption that our data can be represented as a sample from a multivariate Gaussian distribution, the above definition can also be extended to a hold-out target $y_{*}$ at a new location $x_{*}$ in terms of the trained covariance $\mathbf{K}$. The joint probability of training targets $y$ and test targets $y_*$ shown on Eq. \eqref{eq:joint} is also a Gaussian Process:
\begin{equation}\label{eq:joint} 
p(y, y_*) = \mathcal{N}\left( \mathbf{0}, \begin{bmatrix} \mathbf{K} & \mathbf{K}_{*}^{T}\\ \mathbf{K}_{*} & \mathbf{K}_{**} \end{bmatrix}
\right).
\end{equation}

The covariance $\mathbf{K}$ is obtained by $\mathbf{K}_{ij} = k(x_i, x_j)$ is the matrix we get by applying the trained kernel function to our training values, i.e., the similarity of each observed $x$ to each other observed $x$. $\mathbf{K}_{*} = k(x_i, x_*)$ shows the similarity of the training values to the test value whose output values we’re trying to estimate. $\mathbf{K}_{**} = k(x_*, x_*)$ gives the similarity of the test values to each other.

The desired posterior of our prediction is the conditional probability distribution the test target $p(y_*|y)$. This is finally derived using the joint probability distribution in Eq. \eqref{eq:joint} using marginalization:  

\begin{equation}
    p(y_*|y) = \frac{p(y,y_*)}{p(y)} = \mathcal{N}\left( \mathbf{K}_{*} \mathbf{K}^{-1} y, \mathbf{K}_{**} - \mathbf{K}_{*} \mathbf{K}^{-1} \mathbf{K}_{*}^{T}\right).
\end{equation}

Hence the mean of our estimation at new test location is simply given by $\mu(y_*) = \left(\mathbf{K}_{*} \mathbf{K}^{-1} y\right)$, and the uncertainty of this prediction is $\sigma(y_*) = \left(\mathbf{K}_{**} - \mathbf{K}_{*} \mathbf{K}^{-1} \mathbf{K}_{*}^{T}\right)$. In the prediction phase of our emulator, this mean and variance for the predictive weights $w_{i_*}(\theta_{*})$ are calculated for any new set of cosmological parameters $\theta_{*}$. Since the form of $\mathbf{K}$ is already determined from training points using hyperparameter optimization, the GP prediction is simply matrix operations. This makes GP inference extremely fast and easily parallelizable.

%% file: params_appendix.tex
\label{sec:params_appendix}
Table of all parameters of each COLA simulation. 

\clearpage

\setlength\tabcolsep{8pt}

\begin{table*}
\caption{Cosmological parameters of the all the COLA simulations in the suite. 50 training (M00-M49) 
cosmologies are tabulated.}
\sisetup{round-mode=places}
\centering
{
\begin{tabular}{c S[round-precision=3]
S[round-precision=3] S[round-precision=3]
S[round-precision=3] 
S[round-precision=3] }

Model & $\Omega_m h^{2}$ & $n_s$ & $\sigma_{8}$ & log {\it (f$_{R_{0}}$)} & $n$\\
\hline
M00  &   0.14214286 & 1.03775513 &  0.81836736 & -6.36734694 &  1.87755203 \\
M01  &  0.15214285 &  0.91122443 &  0.88367337 & -7.10204082 &  2.28571606 \\
M02  &  0.15357144 &  1.03367341 &  0.777551  & -6.6122449 &  0.81632799 \\
M03  &  0.13499999 &  0.89081639 & 0.82653058 & -4.24489796 &  1.06122398 \\
M04  &  0.14071429 & 0.95204079 &  0.72448981 & -4.97959184 &  1.22448802 \\
M05  &  0.14928573 &  0.96428579 &  0.89591837 & -4.48979592 & 2.04081607 \\
M06  &  0.14571428 &  0.92755097 &  0.76938778 & -7.18367347 &  1.63265204 \\
M07  &  0.13857141 &  0.97653055 &  0.8510204 & -7.26530612 &  0.32653081 \\
M08  &  0.12357143 &  0.96020401 &  0.76530617 & -4.32653061 &  3.42857194\\
M09  &  0.15071428 & 0.87448984 & 0.75714278 & -6.93877551 &  1.38775599\\
M10  &  0.13785714 & 1.0295918 &  0.89183676 & -5.87755102 &  3.34694004\\
M11  &  0.14642857 & 0.9561224 &  0.81428575 & -4.40816327 &  0.0816328 \\
M12  &  0.14285713 & 0.98061216 &  0.87142861 & -4.16326531 &  3.59183598\\
M13  &  0.14357142 & 0.85408169 & 0.74081635 & -6.69387755 &  0.89796001\\
M14  &  0.14142858 &  0.88673461 &  0.84693879 & -6.85714286 &  4.        \\
M15  &  0.1392857 & 0.85      & 0.80612236 & -6.28571429 &  2.44898009\\
M16  &  0.15000001 &  1.04183674 &  0.7204082 & -5.55102041 &  2.53061199\\
M17  &  0.15428573 & 0.894898  &  0.86326539 & -7.34693878 &  1.79592001\\
M18  &  0.14714286 & 1.02142859 &  0.71224487 & -7.91836735 &  3.83673596\\
M19  &  0.13071427 & 0.90306121 & 0.7   & -5.2244898 &  3.51020408\\
M20  &  0.12714286 & 0.98469377 & 0.76122439 & -5.95918367 &  0.73469198\\
M21  &  0.13      &  1.01734698 & 0.81020397 & -7.42857143 &  1.30612397\\
M22  &  0.12      &  0.93163258 &  0.867347  & -6.53061224 &  0.97959202\\
M23  &  0.12214285 &  0.90714282 &  0.79795921 &  -4.   &  0.24489801\\
M24  &  0.14857145 &  1.00102043 & 0.855102  & -6.20408163 &  1.142856  \\
M25  &  0.12642856 &   0.8989796 &  0.74489796 & -5.79591837 &  2.85714412\\
M26  &  0.15285715 &  0.99285716 &  0.80204082 & -8.    &  3.75510406\\
M27  &  0.155     &  0.9683674 & 0.73265296 & -5.46938776 &  3.02040792\\
M28  &  0.14500001 &  0.882653  &  0.78979599 & -4.57142857 & 2.69387603\\
M29  &  0.14785713 &  0.91530621 &  0.83877557 & -6.04081633 &  3.2653079 \\
M30  &  0.14428572 & 0.99693877 &  0.9       & -6.7755102 &  0.48979601\\
M31  &  0.12785715 &  0.87857139 &  0.88775516 & -4.08163265 &  0.1632652 \\
M32  &  0.12071428 & 1.00510192 &  0.73673457 & -4.65306122 &  2.2040801 \\
M33  &  0.12499999 &  0.8581633 &  0.87959176 & -6.44897959 &  1.46938801\\
M34  &  0.13571429 &  1.05      &  0.7938776  & -5.71428571 &  1.55102003\\
M35  &  0.13357142 &  1.01326537 &  0.75306118 & -5.14285714 &  0.        \\
M36  &  0.13142858 &  1.04591835 &  0.85918355 & -5.30612245 &  3.10204005\\
M37  &  0.13214286 &  0.91938782 &  0.72857142 & -7.83673469 &  0.408164  \\
M38  &  0.12857142 &  0.94387758 &  0.78571415 & -7.59183673 &  3.9183681 \\
M39  &  0.13714287 &  0.87040824 & 0.70816326 & -6.12244898 &  0.65306002\\
M40  &  0.12428571 &  1.02551019 &  0.84285718 & -5.06122449 &  2.61224389\\
M41  &  0.15142856 &  0.93979597 &  0.83469379 & -7.67346939 &  0.571428  \\
M42  &  0.13642858 &  0.92346942 &  0.74897957 & -4.73469388 &  3.67346811\\
M43  &  0.12928571 &  0.972449  & 0.83061218 & -4.81632653 &  2.36734796\\
M44  &  0.12571427 &  0.93571419 &  0.70408165 & -7.51020408 &  1.95918405\\
M45  &  0.1342857 &  0.94795918 &  0.87551022 & -7.02040816 &  2.77551198\\
M46  &  0.14000002 &  1.00918353 &  0.77346939 & -7.75510204 &  2.93877602\\
M47  &  0.12142856 &  0.86632657 &  0.82244897 & -5.3877551 &  3.18367195\\
M48  &  0.12285714 &  0.98877555 &  0.71632653 & -4.89795918 &  1.71428394\\
M49  &  0.13285714 &  0.8622449 &  0.7816326 & -5.63265306 &  2.12244797\\
\hline

\end{tabular}%
\npnoround
}
\label{T_Mill}
\end{table*}